\documentclass[]{jfm}

\usepackage{bm}
\usepackage{cancel}
\usepackage{mathtools}
\usepackage{siunitx}
\usepackage{multirow}
\usepackage{graphicx}
\usepackage{subfigure}
\usepackage{newtxtext}
\usepackage{newtxmath}
\usepackage{natbib}

\usepackage{hyperref}
\hypersetup{
colorlinks = true,
urlcolor   = blue,
citecolor  = black,
}

\usepackage[capitalise,noabbrev]{cleveref} 

\title{Space-time nonlinear reduced-order modelling for unsteady flows}

\author{Xiaodong Li\aff{1} \and Davide Lasagna\aff{1}\corresp{\email{davide.lasagna@soton.ac.uk}} }

\affiliation{\aff{1}
Aerodynamics and Flight Mechanics Group, University of Southampton, Boldrewood Campus, Burgess Road, Southampton SO16 7QF, UK
}

\begin{document}
\maketitle

\begin{abstract}
This work investigates projection-based Reduced-Order Models (ROMs) formulated in the frequency domain, employing space-time basis functions constructed with Spectral Proper Orthogonal Decomposition (SPOD) to represent dominant spatio-temporal structures. Although frequency domain formulations are well-suited to capturing time-periodic solutions, such as unstable periodic orbits, this study focuses on modelling statistically stationary flows by computing long-time solutions that approximate the underlying statistics. In contrast to ROMs based solely on spatial modes, a space-time formulation achieves simultaneous reduction in both space and time. This is accomplished by Galerkin projection of the Navier-Stokes equations onto the SPOD basis using a space-time inner product, yielding a quadratic algebraic system of equations in the unknown amplitude coefficients. ROM solutions are obtained by identifying amplitude coefficients that minimise an objective function corresponding to the sum of the squares of the residuals of the algebraic system across all frequencies and modes, quantifying the aggregate violation of momentum conservation within the reduced subspace. A robust gradient-based optimisation algorithm is employed to identify the minima of this objective function. The method is demonstrated for chaotic flow in a two-dimensional lid-driven cavity at $Re=20{,}000$, where solutions with extended temporal periods approximately fifteen times the dominant shear layer time scale are sought. Even without employing closure models to represent the truncated spatio-temporal triadic interactions, multiple ROM solutions are found that successfully reproduce the dominant dynamical flow features and predict the statistical distribution of turbulent quantities with good fidelity, although they tend to overpredict energy at spatio-temporal scales near the truncation boundary.
\end{abstract}

\section{Introduction}

Reduced-order modelling techniques have been developed over several decades to capture coherent structures in fluid flows. These methods offer low-dimensional representations of the complex dynamics of fluid motion and serve as important tools to accelerate numerical prediction and flow control~\citep{holmes_turbulence_2012,rowley_model_2017}. A prevalent approach to building a Reduced-Order Model~(ROM) involves projection methods, in which the velocity field is expanded into a finite set of test basis functions with unknown amplitude coefficients to be determined by solving the ROM. The residual of the governing equations is then made orthogonal to a set of trial functions using a space-only inner product, leading to the Galerkin~\citep{rowley_model_2004} or Petrov-Galerkin methods~\citep{carlberg_efficient_2011}.
Various modal analysis methods have been employed to generate the basis functions, such as Proper Orthogonal Decomposition~(POD), balanced truncation, or linear stability modes~(see the review of \citet{taira_modal_2017}).
The projection operation reduces the high spatial dimensionality of the original problem and yields a ROM in the form of a set of Ordinary Differential Equations~(ODEs).

This space-only modelling framework is particularly effective when the analysis at hand can be recast into the solution of an initial value problem, where the ROM is integrated forward in time from a known initial condition {at a reduced computational cost compared to the full-order system.} Applications include the control of flow instabilities in transitional boundary layers~\citep{semeraro_feedback_2011} and cavity flows~\citep{barbagallo_closed-loop_2009}, {the suppression of turbulence in wall-bounded flows \citep{Maia_Cavalieri_2025}} and the modelling of the transient development of a bluff body wake~\citep{noack_hierarchy_2003} {and its control \citep{Bergmann}}. However, many fluid mechanics problems are better framed through the analysis of the statistical properties of the fully developed turbulent state, i.e.~the statistically stationary state reached when the flow settles onto a chaotic attractor. Such cases include the determination of bounds on time-averaged quantities~\citep{chernyshenko_polynomial_2014}, the development of feedback control strategies for chaotic attractors~\citep{tadmor_reduced-order_2011,lasagna_sum--squares_2016,leclercq_linear_2019}, and  the characterisation of intermodal energy transfers~\citep{couplet_calibrated_2005,rubini_l_2020,jin_energy_2021}.

In such circumstances, space-only ROMs may be inadequate. Firstly, the long-term boundedness of trajectories of a projection-based ROM, and therefore their statistical properties, is challenging to establish~\citep{Schlegel_Noack_2015} or enforce when data-driven model identification methods are used~{\citep{kaptanoglu_promoting_2021, peng}}. Second, these ROMs also unnecessarily retain most of the temporal dimensionality of the original problem, as numerical integration captures a wide range of temporal scales~\citep{choi_space--time_2019,towne_space-time_2021,frame_space-time_2024}. {Third, space-only ROMs exhibit spurious temporal modes, such as exponentially growing or decaying behaviour, which do not pertain to the developed state once the system has settled onto the attractor~\citep{sharma_correspondence_2016}}. Furthermore, closure models~\citep{ahmed_closures_2021} or calibration techniques~\citep{loiseau_constrained_2018,rubini_l_2020,chua_khoo_sparse_2022} are typically required to compensate for the influence of truncated modes on the ROM, particularly in high-Reynolds-number flows, yet constructing such corrections remains a significant obstacle.

\subsection{{The space-time ROM formalism}}

To address these limitations, space-time ROMs have been proposed~\citep{yano_space-time_2014,choi_space--time_2019}. In these approaches, the solution is expanded {into} a set of space-time basis functions and the {unknown} amplitude coefficients are determined {using different techniques from the solution of a linear or nonlinear algebraic problem, depending on the nature of the starting set of partial differential equations}. Unlike space-only ROMs, space-time ROMs directly approximate the complete spatio-temporal trajectory of the system over a predetermined time period, {without requiring temporal integration. Therefore, the desired temporal behaviour and resolution can be controlled a-priori by a suitable choice of the temporal basis.} \citet{choi_space--time_2019} introduced a space-time least-squares Petrov-Galerkin method for nonlinear dynamical systems and applied it to one-dimensional nonlinear advection equations, demonstrating that error bounds grow subquadratically in time, as opposed to the exponential growth typical of space-only ROMs. Applications in optimal control~\citep{baumann_space-time_2018} and hemodynamics~\citep{tenderini_space-time_2024} further illustrate the potential of this approach.

{In the fluid mechanics community, efforts to construct space-time ROMs} have only recently begun to emerge~\citep{towne_space-time_2021,frame_linear_2024,frame_space-time_2024,li_nonlinear_2024}
{ using resolvent modes~\citep{jovanovic_componentwise_2005,mckeon_critical-layer_2010,sipp_dynamics_2010} or spectral POD (SPOD) \citep{PICARD2000359, towne_spectral_2018}. }
{\citet{towne_space-time_2021} utilised these two bases} in Galerkin and Petrov-Galerkin approaches, and found that space-time SPOD Petrov-Galerkin ROMs can produce accurate solutions with low computational cost in a one-dimensional~(1D) linearised Ginzburg-Landau problem. In \citet{frame_linear_2024}, applications to a 1D linearised Ginzburg-Landau problem and a 2D linear advection-diffusion problem have shown error reductions {by} several orders of magnitude compared to space-only methods, e.g.~POD-Galerkin and balanced truncation, along with improved computational efficiency. Similar conclusions were reported in a subsequent study by \citet{frame_space-time_2024}, where the approach was extended to 1D nonlinear Ginzburg-Landau systems that incorporate standard cubic and Burgers-type nonlinearities{, using an iterative fixed-point technique to solve the resulting nonlinear algebraic system governing the amplitude coefficients of the space-time expansion.} As noted in \citet{frame_space-time_2024}, this method may not always converge robustly, particularly under conditions of strong nonlinearity, although the circumstances influencing convergence remain to be fully understood and may vary with problem characteristics, especially in high-Reynolds-number flow regimes. 

\subsection{{Relation with harmonic balance methods and unstable periodic orbits}}
{When formulated using a Fourier basis in time, the case of resolvent and SPOD modes, space-time ROM techniques are closely related to harmonic balance ~\citep{he_efficient_1998, hall_computation_2002, hall_harmonic_2013}. These methods have been popular in the turbomachinery community, where the temporal periodicity is inherent in the rotation of the system under study, but have also been applied more recently to transitional boundary layer flows \citep{Rigas_Sipp_Colonius_2021}, where this is not the case, and to capture limit cycle oscillations at moderate Reynolds numbers~\citep{SIERRAAUSIN2022114736}, where the oscillation period is obtained from the solution process. In these works, a system of nonlinear algebraic equations containing the unknown Fourier components of the time-periodic velocity field is derived and solved using root-finding techniques. Space-time modelling methods make a further step by expanding each Fourier component into a set of orthonormal spatial basis functions weighted by unknown amplitude coefficients, to reduce the dimensionality of the problem and achieve computational savings by solving a smaller system of algebraic equations.}

{For turbulent flows with chaotic dynamics, a continuous spectrum and lacking strong periodic behaviour, the use of a temporal Fourier basis may still be justified in the limit of infinite time periods by the temporal homogeneity of flow statistics, akin to a large periodic spatial domain being used to approximate flows with spatial translational invariance (e.g. channels or pipes) \citep{sharma_correspondence_2016}. This constitutes the most natural function space in which to describe the dynamics once the system has settled onto its attractor. More importantly, for finite (but unknown) time periods, the Fourier basis is still justified as it is the natural representation of Unstable Periodic Orbits (UPOs), i.e.~time-periodic exact solutions of the Navier-Stokes equations (see \citet{kawahara_significance_2012} and \citet{graham_exact_2021} for reviews). This property has motivated recent developments in UPO search methods \citep{parker_variational_2022, Burton2025ResolventBased}, which utilise a Fourier-in-time representation of the time-periodic velocity field.}

{The relationship with UPOs and the dynamical system theory viewpoint of turbulence needs to be further discussed. In principle, detailed knowledge of the hierarchy of short-period UPOs is sufficient to estimate infinite-time averages using cycle expansion theory ~\citep{artuso1990recycling}. However, locating UPOs in fluid systems is challenging \citep{page_recurrent_2024}, posing hurdles to the application of cycle expansion theory to turbulence \citep{Chandler_Kerswell_2013, yalniz_coarse_2021, Wang_Ayats_Deguchi_Meseguer_Mellibovsky_2025}. This issue is relevant, since the quality of cycle expansion theory predictions using an incomplete hierarchy of UPOs is as good as the most important orbit that one fails to locate \citep{Budanur2015}. In light of such issues, previous work \citep{ lasagna_sensitivity_2020} proposed a heuristic approach whereby available computational resources are spent to locate one or a few UPOs having a sufficiently long period $T$. Using low-dimensional systems, such as the Lorenz equations, it was shown that time-averaged quantities and probability distributions computed on such UPOs appear to converge to those computed from chaotic trajectories as $T$ increases. This is intuitively justified by the fact that these orbits span the invariant measure of the chaotic attractor and provide a scaffold for chaotic trajectories to develop. More importantly for the purpose of this work, it was also shown that long UPOs provide access, via a well-behaved continuation analysis, to the sensitivity of time-averaged quantities with respect to problem parameters \citep{lasagna_sensitivity_2018}, as the periodicity effectively suppresses the exponential growth of the adjoint field \citep{wang_forward_2013}, which makes classical adjoint sensitivity methods fail for chaotic systems \citep{lea_sensitivity_2000}. 
Understanding how the statistics of a turbulent flow depend on problem parameters (such as feedback control gains, geometry, wall attributes, design parameters, etc) may be a key enabler to develop novel turbulence control strategies, but this information is hardly accessible in practice. }

\subsection{{Motivation of the present study and main contributions}}
{Given that computing long period UPOs remains challenging, the tight mathematical and algorithmic connections between UPOs and space-time ROMs indicate that the latter may present an efficient and promising alternative. Solutions of the space-time ROM may be used to describe at a reduced cost the long-time statistics of turbulent quantities pertaining to the full-order system. Then, when control and design parameters appear directly in the ROM, a numerical continuation study on ROM solutions, or using adjoint sensitivity methods, may shed light on how these statistics depend on such parameters. The adjoint sensitivity problem would be formulated directly on the reduced subspace as in \citet{karbasian_sensitivity_2022}, rather than in the complete state space, as in \citet{Citro}, further reducing computational costs. Nevertheless, there are several fundamental hurdles in this programme of work that need addressing before this framework could be utilised for flow control and optimisation. We name a few key ones. First, nonlinear space-time methods needs to be extended to and tested on fluid systems governed by the Navier-Stokes equations. Second, the relation between time-periodic solutions of a ROM and UPOs needs to be investigated in depth. Advances on this front have already been made \citep{mccormack_multi-scale_2024}. Third, for sensitivity analysis, the space-time basis needs to be sufficiently rich to capture a potentially complex parametric dependence, although progress may be made by leveraging model order reduction methods designed for time-periodic fluid systems, as in \citet{PADOVAN2024112597}.}

{In this paper, we only focus on addressing the first of these hurdles, and develop and test a nonlinear space-time reduced-order modelling method for fluid systems governed by the incompressible Navier-Stokes equations. We employ a data-driven approach and use SPOD modes as the space-time basis. Similar to \citet{frame_space-time_2024}, Galerkin projection of the Navier-Stokes equation on this basis produces a nonlinear algebraic system, the ROM, governing the amplitude coefficients of the velocity expansion. In contrast to that work, we propose to use gradient-based methods to compute solutions of the ROM robustly, by finding minima of an objective function corresponding to the average violation of momentum conservation in the reduced subspace. The approach is demonstrated on two-dimensional lid-driven cavity flow at a Reynolds number $Re=20{,}000$, where the dynamics exhibit chaotic behaviour. We illustrate the ability of the ROM to recover low-dimensional structures embedded in the high-dimensional chaotic dynamics, producing a statistical distribution of physical quantities in good agreement with that of DNS, despite no closure of calibration techniques being used.}

The remainder of this paper is organised as follows. In \cref{sec:methodology}, we formulate the nonlinear {space-time ROM} and describe the optimisation strategy to find its solutions. \cref{sec:flow_SPOD} presents the numerical discretization and SPOD analysis of the lid-driven cavity flow. The optimisation-based approach to solve the ROM is detailed in \cref{sec:ROM_optimisation}, and results on flow dynamics and statistics are discussed in \cref{sec:APO_results}. {Finally, \cref{sec:conclusions} summarises the key findings and presents future directions.}

\section{Methodology}\label{sec:methodology}

\subsection{Space-time reduced-order modelling}\label{sec:methodology_ROM}
Unsteady, incompressible flows governed by the nondimensional continuity equation and Navier--Stokes equations,
{\begin{equation}
\begin{aligned}
    \nabla \cdot \bm{u} &= 0\\
    \frac{\partial  \bm{u}}{\partial t} + (\bm{u} \cdot \nabla )   \bm{u} -\frac{1}{Re} \nabla \cdot \nabla  \bm{u} &= -\nabla p
    \, ,
\end{aligned}
\label{equ:NS_incomp}
\end{equation}}
are considered, where $\bm{u}$ and $p$ denote velocity and pressure. {To model the developed turbulent state that establishes once the flow has settled onto the attractor, we consider a time-periodic velocity field defined over a finite period $T$ that is long enough so that finite-time averages computed over this field may be considered good approximations of infinite-time averages~\citep{lasagna_sensitivity_2020}. Such a field may be adequately approximated by an expansion into separable space-time basis functions that rely on the Fourier basis to describe the temporal evolution~\citep{lumley_stochastic_1970, sharma_correspondence_2016}, automatically enforcing periodicity, and appropriate spatial basis functions to capture the coherent structure of turbulence at each frequency.} An expansion of the velocity field over such an interval may thus be
\begin{equation}
\begin{aligned}
\bm{u}(\bm{x}, t) &= {\bm{U}(\bm{x})} + \bm{u}'(\bm{x},t) \\
&= {\bm{U}(\bm{x})} + \bm{u}'_{r}(\bm{x},t) + \bm{u}'_{u}(\bm{x},t)\\
&= {\bm{U}(\bm{x})} + \sum_{k=-N}^{N} \sum _{j=1}^{M} a_{j}^{k} \bm{\phi}_{j}^{k}(\bm{x}) e^{i k \omega t}
+ \underset{ {\mathclap{\scriptscriptstyle \max \left( |k|-N, j-M \right)>0}} }{ \sum_{k=-\infty}^{\infty} \sum_{j=1}^{\infty} \strut } a_{j}^{k} \bm{\phi}_{j}^{k}(\bm{x}) e^{i k \omega t}
\, ,
\end{aligned}
\label{equ:SPOD_finite_truncated_mean}
\end{equation}
where the infinite-time-averaged flow and the deviation field are denoted by ${\bm{U}} $ and $ \bm{u}'$, respectively. The latter is composed of resolved motions $\bm{u}'_{r}(\bm{x},t) $ and unresolved motions $\bm{u}'_{u}(\bm{x},t) $ at frequencies that are integer multiples of the fundamental frequency $\omega = 2 \pi / T$.
Note that the deviation from the infinite-time mean includes motions for $k=0$, which account for the possibility that the mean computed over the finite-time solution may be different from the infinite-time mean, even if this difference may become smaller for large $T$. This choice follows the {empirical} evidence that period averages computed on UPOs of chaotic systems are all different, although such differences converge to zero when long-period UPOs are considered~\citep{saiki_time-averaged_2009}. The superscript $k$, bounded in the range $[-N, N]$, denotes the index of discrete frequencies, so that the resolved frequencies are $f^k = k/T$. On the other hand, the subscript $j$ is used as the index of the basis functions at each frequency. The truncated space-time modes, for $|k| > N$ or $j > M$ are lumped into the unresolved component $\bm{u}'_{u}(\bm{x}, t) $. We assume that any inhomogeneous boundary condition is handled by the infinite time mean $\bm{U}(\bm{x})$. Then, the spatial modes $\bm{\phi}_{j}^{k}(\bm{x})$ only need to satisfy homogeneous boundary conditions on all boundaries of the domain. The associated amplitude coefficients, denoted as $a_j^k$, are the unknown variables to be determined. Because velocity fields are real valued, the amplitude coefficients (and the spatial modes) satisfy the conjugate symmetry, i.e., $a_j^{-k} = \overline{a_j^{k}}$, with $ \overline{( \cdot )}$ denoting the complex conjugate operation.

In what follows, we make use of the space-only inner product
\begin{equation}
\left( \bm{u}, \bm{v} \right) = \int_{V} \overline{\bm{u}} \cdot \bm{v} \, \mathrm{d}V
\end{equation}
between two generic vector fields $\bm{u}(\bm{x})$ and $\bm{v}(\bm{x})$, which only includes integration over the spatial domain $V$ and induces the norm $\|\bm{u}\| = \sqrt{\left ( \bm{u}, \bm{u}\right )}$. We also use the space-time inner product
\begin{equation}
\left [ \bm{u}, \bm{v} \right ] = \int_{0}^{T} \int_{V} \overline{ \bm{u} } \cdot \bm{v} \, \mathrm{d}V \, \mathrm{d}t\,,
\label{equ:space_time_inner_product}
\end{equation}
between two space-time vector fields $\bm{u}(\bm{x}, t)$ and $\bm{v}(\bm{x}, t)$, which also includes integration over the temporal domain.

We follow a data-driven approach and consider SPOD modes~\citep{PICARD2000359, towne_spectral_2018} to construct the space-time basis. Other choices may be considered, such as the modes obtained from resolvent analysis, as discussed in \citet{towne_space-time_2021}. {Note that the completeness of the basis may have consequences on whether expansion \eqref{equ:SPOD_finite_truncated_mean} may represent, upon refinement, exact solutions of the equations, i.e.~UPOs.} Then, Galerkin projection is employed to achieve momentum conservation in the low-order subspace. The frequency domain ROM is obtained by substituting the velocity expansion of equation \eqref{equ:SPOD_finite_truncated_mean} in the governing equations \eqref{equ:NS_incomp} and then projecting them onto each of the space-time basis functions $\bm{\phi}_{m}^{l}(\bm{x}) \, e^{i l \omega t}$ in turn, using the space-time inner product. This operation yields a system of nonlinear algebraic equations in the low-order subspace
\begin{equation}
\begin{aligned}
r^l_m \coloneq & \sum_{n=1}^{M} a_{n}^{l} \left( i l \omega \delta_{m,n} + L^{l,l}_{m,n} \right) 
+ \sum_{n=1}^{M} \sum_{p=1}^{M} \sum_{k=-N}^{N} a_{n}^{k} a_{p}^{l-k} Q^{k,l-k,l}_{m,n,p} + C^l_m \delta_{l,0} + G_m^l + {P_m^l}
= 0
\, ,
\end{aligned}
\label{equ:NS_incomp_SPOD_proj_model}
\end{equation}
for $l\in [-N, N]$, $m\in [1, M]$, with $ \delta_{l,0}$ being the Kronecker delta. Model coefficients in \cref{equ:NS_incomp_SPOD_proj_model} are represented by the tensors $\bm{L}, \bm{Q}, \bm{C}$ and are determined from the space-only inner product, owing to the orthogonality of Fourier modes, between sets of SPOD modes and the infinite-time-averaged velocity field as
\begin{equation}
\begin{aligned}
L^{l,l}_{m,n} &= \left( \bm{\phi}_{m}^{l} , \bm{U} \cdot  \nabla \bm{\phi}_{n}^{l} \right) + \left( \bm{\phi}_{m}^{l} , \bm{\phi}_{n}^{l} \cdot \nabla \bm{U} \right) - \frac{ 1 }{Re}  \Big( \bm{\phi}_{m}^{l} ,  \nabla \cdot \nabla \bm{\phi}_{n}^{l} \Big), \\
Q^{k,l-k, l}_{m,n,p} &= \Big( \bm{\phi}_{m}^{l} , \nabla \cdot ( \bm{\phi}_{n}^{k} \bm{\phi}_{p}^{l-k} ) \Big), \\
C^{l}_m &= -\frac{1}{Re} \Big( \bm{\phi}_{m}^{l} , \nabla \cdot \nabla \bm{U}  \Big) 
+ \left(  \bm{\phi}_{m}^{l} , \bm{U} \cdot \nabla \bm{U} \right)
\, .
\end{aligned}
\label{equ:model_coeffs_expression}
\end{equation}
{The derivations of these definitions are reported in  \cref{app:model_coeffs}.} The tensor $\bm{C}$ denotes constant model coefficients arising from the convective {and viscous} terms of the governing equations and the infinite-time-averaged velocity field. The tensor $\bm{L}$ contains model coefficients arising from linear mechanisms in the governing equations, such as advection with the mean field and viscous diffusion, while the tensor $\bm{Q}$ denotes model coefficients associated with fluctuation-fluctuation nonlinearity. All these coefficients represent the interaction between sets of SPOD modes at different frequencies and with the infinite-time-averaged velocity field, as summarised in \cref{tab:model_coeff}.
\begin{table}
\begin{center}
\def~{\hphantom{0}}
\begin{tabular}{lccc}
                                    & $\bm{L}$ & $\bm{Q}$ & $\bm{C}$ \\[3pt]
Reliance on infinite time averaged field         & Yes & --  & Yes \\%
Modes coupled at each frequency                  & Yes & Yes & --  \\
Modes coupled across frequencies                 & --  & Yes & --  \\
\end{tabular}
\caption{Properties of model coefficients in the {space-time} ROM.}
\label{tab:model_coeff}
\end{center}
\end{table}
Given that SPOD modes satisfy the conjugate symmetry, the model coefficients obey
\begin{equation}
L^{-l,-l}_{m,n} = \overline{L^{l,l}_{m,n}} \quad \mathrm{and}\quad Q^{-k,-l+k, -l}_{m,n,p} = \overline{Q^{k,l-k,l}_{m,n,p}}.   
\end{equation}
Furthermore, because of the convolution sum over different frequencies and modes that involve the quadratic coefficients $\bm{Q}$, the frequency domain ROM proposed here is fully nonlinear.
The {tensor $\bm{G}$} represents the impact of truncated modes on the nonlinear ROM, and consists of unresolved triadic interactions with the mean flow field, resolved fluctuations, and unresolved fluctuations, as presented in \cref{app:unresolved_interaction}. In principle, closure models could be developed that express the {tensor $\bm{G}$} as a function of resolved spatio-temporal scales, e.g.~using classical eddy viscosity models (see \citet{ahmed_closures_2021}) specialised to the present frequency domain setting. Here, we investigate the Galerkin model without considering the unresolved interactions, viz.~$G_m^l=0$, and leave the development of closure techniques to future work.
{The tensor $\bm{P}$ includes the contribution related to the pressure gradient term projected onto the space-time basis functions.
This term vanishes for flows with certain boundary conditions, such as no-slip or periodic conditions, like in the lid-driven cavity flow considered in this paper, as illustrated in \cref{app:pressure_gradient}.}

{An important remark is that the choice of the inner product for the projection differs from classical POD-Galerkin methods for model order reduction, where the space-only inner product is utilised to derive a space-only ROM consisting of a set of ODEs that yield, upon time-marching, the temporal coefficients of the velocity expansion. 
In the present framework, we use the space-time inner product \cref{equ:space_time_inner_product} and constrain the space of admissible solutions to time-periodic velocity fields, which constitutes the most natural space in which to describe the dynamics once the system has settled onto its attractor. Solving the residual equations \eqref{equ:NS_incomp_SPOD_proj_model} is, effectively, akin to solving a nonlinear periodic boundary value problem using the Fourier-Galerkin method on a reduced subspace.
%
%
The residual equations still identify a reduced-order model, except that solutions of the ROM now define low-dimensional representations of entire space–time trajectories.} {By construction, these trajectories cannot exhibit transient phenomena, such as relaminarisation or temporal blow-up, as often observed in POD–Galerkin models. Nevertheless, the solutions may still display non-physical behaviour, albeit of a different kind. For instance, the ROM may fail to admit any solution, which likely indicates that trajectories of a space-only ROM constructed using the same spatial basis functions would diverge to infinity. Even when solutions exist, they may occupy regions of state space far from those explored by DNS trajectories, signalling an incorrect energy balance within the ROM. Therefore, checking whether the computed solutions are located in the vicinity of the turbulent attractor and reproduce the DNS statistics with good fidelity becomes necessary.} 


\subsection{Optimisation-based solution of the ROM}\label{sec:grad_optim}
In practice, the solutions of the ROM are found by looking for amplitude coefficients for which the residual $r^l_m$ of the system \eqref{equ:NS_incomp_SPOD_proj_model} vanishes simultaneously for all $l$ and $m$. The hypothesis is that if the fundamental period $T$ is long enough, these solutions should provide a good approximation of the statistically steady state observed in DNS.

A solution of the algebraic system \eqref{equ:NS_incomp_SPOD_proj_model} may, in principle, be obtained by using the Newton-Raphson method{, commonly used in harmonic balance techniques}. However, in the initial phases of the research, it was not clear whether the modal truncation might have implied the absence of solutions or even if the high dimensionality and non-linearity of the problem would permit solutions to be found readily with this method, which is known to be sensitive to poor initial conditions. In addition, recent work on search methods for invariant solutions of the Navier-Stokes equations \citep{farazmand_adjoint-based_2016, ashtari_identifying_2023} has adopted adjoint-based variational approaches, whereby the search is recast as the minimisation of a cost function that involves the overall violation of the governing equations. \citet{Barthel_Zhu_McKeon_2021} and then later \citet{burton_resolvent-based_2025} further extended this optimisation-based approach to low-dimensional settings. The approach has been shown to be quite robust to poor initial guesses and it was therefore adopted here. Hence, we consider here the least-squares optimisation problem
\begin{equation}
\begin{aligned}
\min_{ \displaystyle \omega, \{a_j^k\}^{k \in [0, N]}_{n \in [1, M]} } \quad & J = \sum_{m=1}^{M} \sum_{l=-N}^{N} \left| r^l_m \right|^2
\,
\end{aligned}
\label{equ:NS_incomp_SPOD_proj_model_res_obj}
\end{equation}
defined by a non-negative objective function. When a minimum of the objective function is found for which $J=0$, then clearly $r_m^l=0$ for all $m$ and $l$, resulting in a ROM solution that conserves momentum in the low order subspace. 

An important remark is that, when searching for UPOs of the Navier-Stokes equations, the period $T$ is often not known a priori but is found during the search. The fundamental frequency $\omega$ plays the same role here, as it appears directly in the algebraic system \eqref{equ:NS_incomp_SPOD_proj_model} as an unknown. Hence, in addition to the amplitude coefficients, this variable should also be adjusted. However, when obtaining the space-time basis functions with SPOD, the fundamental frequency is determined by the length of the blocks of data used. Allowing the fundamental frequency to vary during the optimisation would break the consistency between the ROM and the modes. Nevertheless, in practice, we find that the relative change of the fundamental frequency during the optimisation is quite small. We thus neglect this inconsistency, as articulated in \cref{sec:ROM_optimisation}, and later show including $\omega$ in the optimisation step is necessary to find ROM solutions.

A further remark is that because of the conjugate symmetry property of the amplitude coefficients, the number of optimisation variables can be reduced by nearly a factor of two by only optimising over the coefficients corresponding to non-negative frequencies, which lifts the requirement of using constrained-optimisation techniques. Overall, the total number of optimisation variables is equal to $M(2N+1) + 1$, including the fundamental frequency.

Gradient-based methods are used to efficiently and robustly find solutions to the optimisation problem \eqref{equ:NS_incomp_SPOD_proj_model_res_obj}. The gradient of the objective function with respect to the amplitude coefficients can be derived analytically as
\begin{equation}
\begin{aligned}
\frac{ \partial J }{\partial {a_e^q} }
& = \sum_{m=1}^{M} \sum_{l=-N}^{N} \left[  \overline{r^l_m}  \frac{ \partial  r^l_m  }{\partial {a_e^q}} +   r^l_m \frac{ \partial \overline{r^l_m}  }{\partial {a_e^q}} \right]\\
&= \sum_{m=1}^{M} \sum_{l=-N}^{N} \left[ \overline{r^l_m} \left(\delta_{q,l} \left( i l \omega \delta_{m, e} +  L^{l, l}_{m, e} \right) + \sum_{n=1}^{M} a_{n}^{l-q} ( Q^{q,l-q, l}_{m,e,n} + Q^{l-q,q,l}_{m,n,e} ) \right) \right. \\
& \left. + r^l_m \left( \delta_{l,-q} \left( -i l \omega \delta_{m, e} +   \overline{L^{l, l}_{m, e}} \right) +  \sum_{n=1}^{M} \overline{a_{n}^{l+q}} ( \overline{{Q}^{-q,l+q, l}_{m,e,n}} + \overline{{Q}^{l+q,-q, l}_{m,n,e}} )  \right) \right]
\, ,
\end{aligned} 
\label{equ:SPOD_model_min_obj_gradient_expession}
\end{equation}
where the subscript $e \in [1, M]$ and the superscript $q \in [-N, N]$ are introduced here to represent the index of modes and frequencies, avoiding the ambiguity of those used in $r_m^l$. Likewise, the analytical expression for the gradient of the objective function with respect to $\omega$ is
\begin{equation}
\begin{aligned}
\frac{\partial J}{\partial \omega} &= 
\sum_{m=1}^{M} \sum_{l=-N}^{N} \left( \frac{\partial \overline{r_m^l}  }{\partial \omega} r_m^l + \overline{r_m^l} \frac{\partial   r_m^l}{\partial \omega} \right)
= \sum_{m=1}^{M} \sum_{l=-N}^{N} 2 l \Im \left(\overline{a_m^l} r_{m}^{l} \right)
\, ,
\end{aligned}
\label{equ:grad_omega}
\end{equation}
where $ \Im( \cdot) $ denotes the imaginary component of a complex number. We use limited-memory BFGS~(L-BFGS) to solve the optimisation problem, as it demonstrated faster convergence rates compared to other simpler gradient descent methods. The optimisation variables are complex-valued, but they are mapped to real values to enable using the optimisers available in the publicly available Julia library \textit{Optim.jl}~\citep{mogensen_optim_2018}. In practice, convergence is deemed sufficient when the norm of the gradient $\| \nabla J \|$ falls below a specified tolerance of $10^{-12}$, ensuring a high level of numerical accuracy. 

The objective function is a quartic polynomial in terms of each amplitude coefficient, and {may thus} admit multiple minima, corresponding to multiple ROM solutions. This complexity reflects the dynamical point of view of turbulence, interpreted as a trajectory in a high-dimensional space that transits through the neighbourhoods of a large number of unstable invariant solutions, including UPOs~\citep{hopf_mathematical_1948,kawahara_periodic_2001,hof_experimental_2004,suri_capturing_2020}.  {Here, minima that reach $J=0$, provided these exist, correspond to actual ROM's solutions as all residuals $r_m^l$ can be reduced arbitrarily close to zero to satisfy momentum conservation. These are referred to as zero-minima in what follows. In contrast, it may happen that the optimisation is convergent but the objective function $J$, and therefore the residuals $r_m^l$, do not converge to zero. These are not solutions of the ROM and we refer to them in what follows as non-zero minima. These are entirely analogous to the concept of ghost states recently proposed by \citet{zheng2024ghost} for nonlinear dynamical systems, i.e.~states that minimise a particular non-negative function that reaches exactly zero only for invariant solutions of the system. Ghost states arise when invariant solutions disappear after a bifurcation, but their dynamical influence persists much after and leads to slow evolution in the case of bifurcating equilibria and near-periodic behaviour in the case of bifurcating periodic orbits. To distinguish zero and non-zero minima, a heuristic metric}
\begin{equation}
\begin{aligned}
\eta = J / \| \nabla J \|\,,
\end{aligned}
\end{equation}
is used and examined upon convergence. Since the objective function can be approximated by a quadratic function near {a minimum}, it is easy to show that $\eta$ will converge to zero when the optimiser converges to a {zero-minimum, but will diverge when the optimiser converges to a non-zero one.}

\section{Application to lid-driven cavity flow}\label{sec:flow_SPOD}
The proposed approach is now investigated for two-dimensional, lid-driven cavity flow in a square cavity, an extensively used test bed for the development of modelling techniques~\citep{cazemier_proper_1998,balajewicz_low-dimensional_2013,rubini_l_2020}. We consider the chaotic flow regime establishing at $Re=20{,}000$, with the top lid moving at uniform velocity and the other three boundaries remaining stationary. The length of the cavity and the lid velocity are used to non-dimensionalise all quantities. 

\subsection{Numerical simulation of lid-driven cavity flow}
Direct Numerical Simulation~(DNS) of this flow was performed using OpenFOAM. The velocity and pressure are solved with Pressure-Implicit with Splitting of Operators~(PISO) algorithm using a dimensionless time step of $0.005$. A second-order central difference method is applied for spatial discretisation, and a second-order Euler backward scheme is used for time direction. Linear interpolation is used to compute the flux and variables at the cell surfaces. The absolute and relative tolerances for convergence of numerical solutions are set to $10^{-5}$.
The computational domain is discretised using a stretched mesh with $39600$ quadrilateral grid cells, which resolves all relevant scales of motion~\citep{rubini_l_2020}. 

\begin{figure}
\centering
\includegraphics[width=\textwidth]{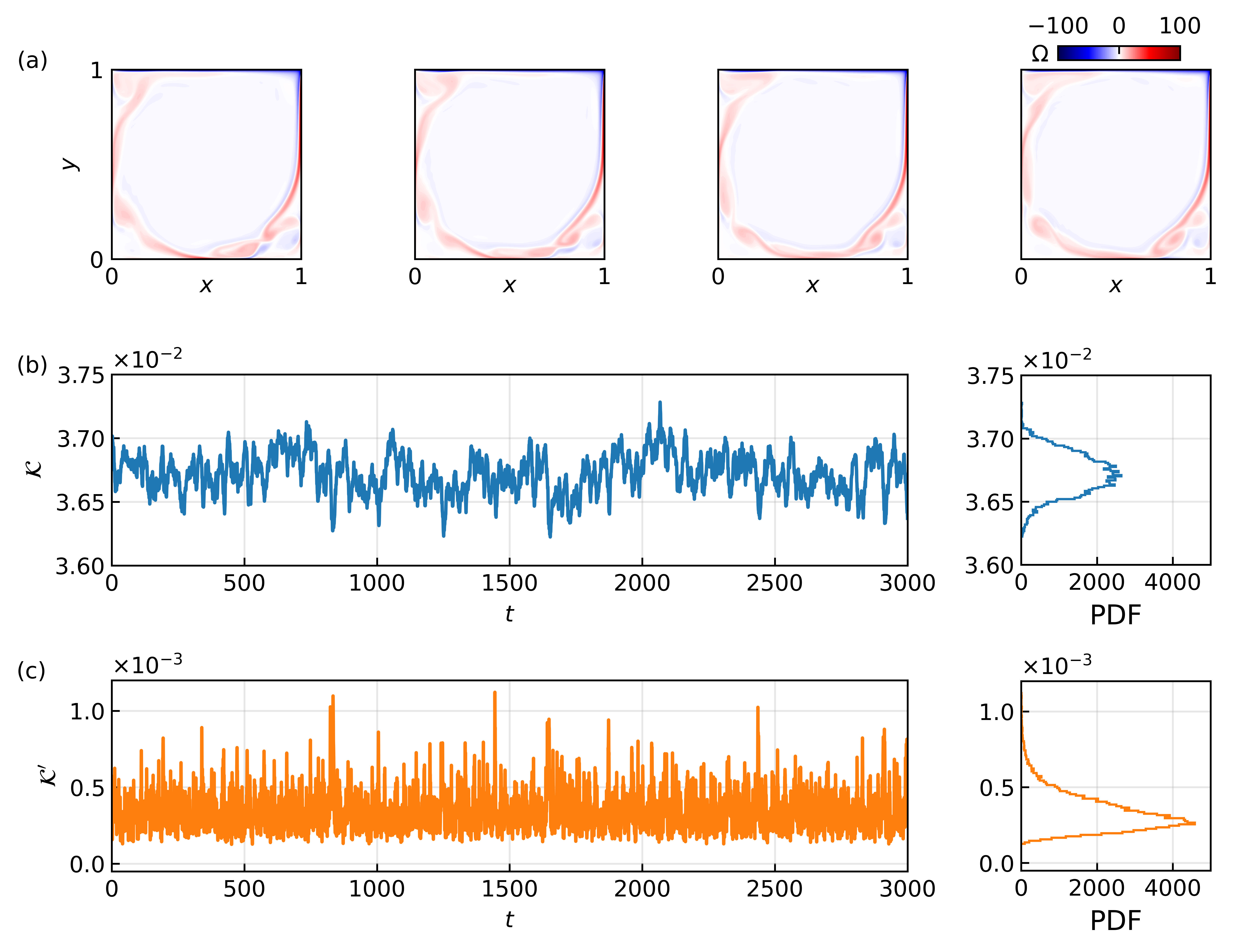}
\caption{Four snapshots of the vorticity field of the fully-developed state separated by a time interval $\Delta t = 0.4$, panel (a). Time history of kinetic energy~($\mathcal{K}$) and turbulent kinetic energy~($\mathcal{K}'$) with the corresponding Probability Density Function~(PDF), panels (b-c).}
\label{fig:vorticity_timeSeries}
\end{figure}

Four instantaneous snapshots of the out-of-plane vorticity {$\Omega(\bm{x}, t)$} sampled with a time interval of $\Delta t=0.4$ from the fully developed state are shown in \cref{fig:vorticity_timeSeries}(a).
The dominant unsteady motions take place primarily in the shear layer bounding the central vortex. Spectral analysis of velocity time series in these regions suggests that the time scale of this oscillation is $T_{\mathrm{osc}} \approx 1.7$. This shear layer interacts with smaller counter-rotating vortices in the corners, where the shear layer is perturbed, leading to nonlinear dynamics inside the cavity.  Highly unsteady flow is observed in the bottom-right corner compared to the other corners.

\cref{fig:vorticity_timeSeries}(b) and (c) show the time history and the Probability Density Function~(PDF) of the instantaneous kinetic energy and the turbulent kinetic energy, defined as
\begin{eqnarray}
\mathcal{K}(t)  &&= \frac{1}{2} \| \bm{u}(\bm{x}, t)  \|^2 \, ,  \quad
\mathcal{K}'(t) = \frac{1}{2} \| \bm{u}'(\bm{x}, t) \|^2
\, ,
\end{eqnarray}
respectively, over a long simulation run for $5000$ non-dimensional time units. {The initial transient over the first 2000 time units, during which the kinetic energy gradually increases towards a range pertaining to the developed state}, has been discarded. The remaining data, shown in the figure, suggest that a statistically steady state has been reached, even though low-frequency motions, with a time scale of hundreds of time units, can be observed. The PDF of the kinetic energy exhibits a Gaussian-like distribution, but for the turbulent kinetic energy the distribution is skewed and exhibits a heavy tail, associated with the bursting of the secondary vortical flow in the bottom-right corner.

\subsection{SPOD analysis}
Ten thousand snapshots are sampled from DNS at a frequency $f_{\mathrm{s}}=10${, from the first 1000 time units of the available data}. To perform SPOD, we select a period $T = 25.6$, leading to $N_f=256$ distinct frequencies. This period is long enough to cover about $15$ cycles of the fundamental shear layer motion and to capture flow dynamics over a wide range of scales at reasonable cost of computation. These settings allow us to resolve frequencies from $\Delta f = T^{-1} = 0.0391$ up to the Nyquist frequency $f_{\mathrm{s}}/2 = 5$. For the Welch method, we select an overlap of 50\%, leading to $77$ data blocks. The {infinite-time-averaged} field {${\bm{U}}(\bm{x})$ is computed from the available} snapshots and is then subtracted from the data prior to analysis.

\begin{figure}
\centering
\includegraphics[width=\textwidth]{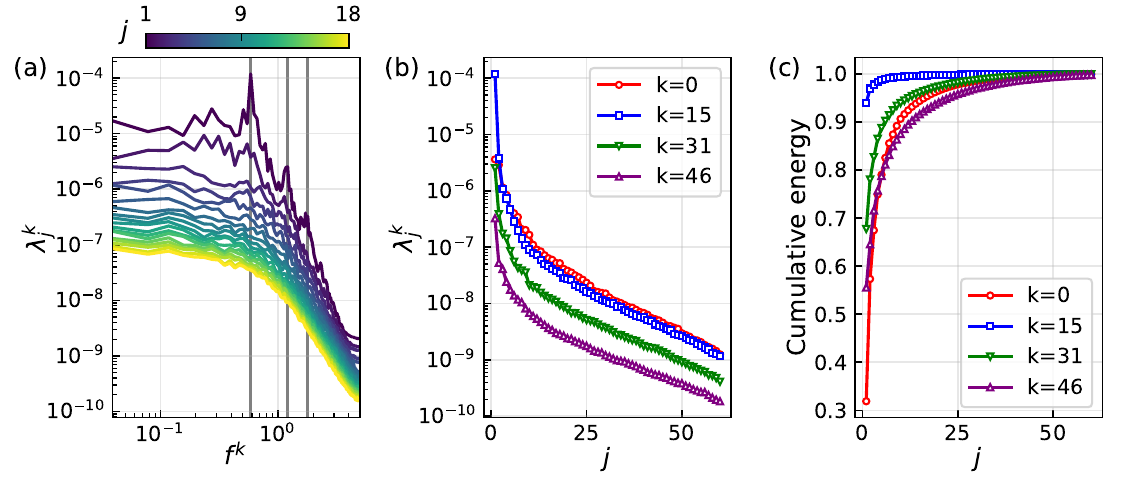}
\caption{Spectrum of SPOD eigenvalues, {up to mode 18, coloured by the mode index, panel (a)}. The vertical lines identify energy peaks at $f^k= 0.586, 1.211, 1.797$. {The decay of eigenvalues for the mean component and these three energy-peak frequencies and their cumulative energy are shown in panels (b) and (c).}}
\label{fig:SPOD_spectra_truncation}
\end{figure}

\cref{fig:SPOD_spectra_truncation} shows {details} of the SPOD eigenvalue spectrum, where $\lambda_j^k$ denotes the $j$-th eigenvalue at each discrete wavenumber $k$. The most energetic mode is found at $f^{15} = 0.586$, which is close to the dominant frequency $T_{\mathrm{osc}}^{-1}=0.588$ identified earlier. {The dynamics at this frequency is of relatively low rank, as the first two modes capture about 95\% of the energy (panel c), while motions at other frequencies require a significantly higher number of modes to reach the same level.} For the first SPOD mode, peaks at the second and third harmonic frequencies are observed (shown by the vertical lines in the figure). In addition, a significant amount of energy is still contained at low frequencies related to motions occurring in the cavity over long time scales, as discussed earlier. It should be noted that the spectrum flattens at high frequencies (i.e., for \( f^k > 3 \)), since the data blocks are not windowed prior to the Fourier transform. These high-frequency components are ultimately excluded from the reduced-order model after truncation.

\begin{figure}
\centering
\includegraphics[width=\textwidth]{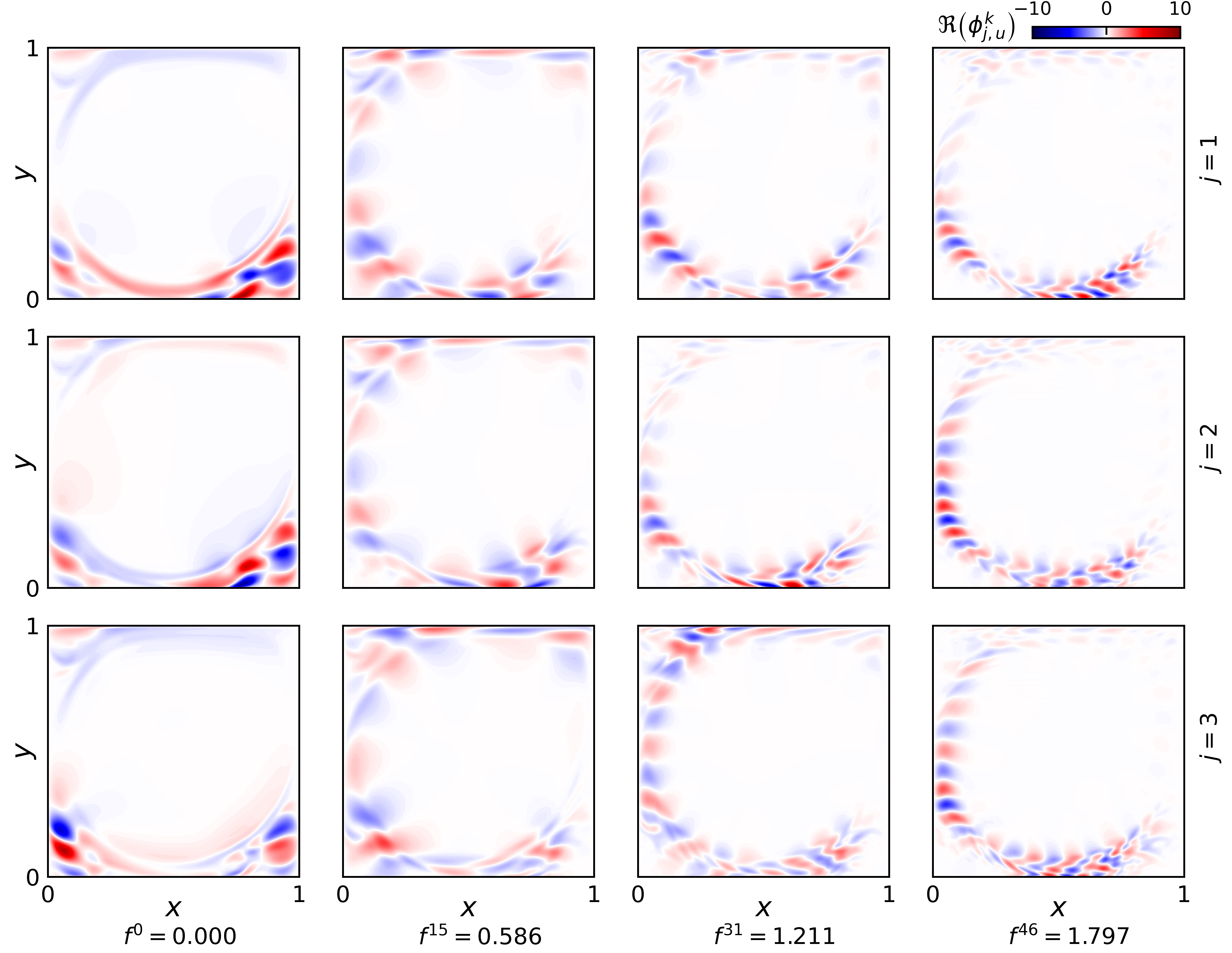}
\caption{Real part of the $x$-component of the first three SPOD modes for the mean component and the three peak frequencies marked by vertical lines in \cref{fig:SPOD_spectra_truncation}(a).}
\label{fig:SPOD_modes}
\end{figure}

The real part of the $x$-component of the SPOD modes for five selected frequencies is shown in \cref{fig:SPOD_modes} to illustrate the spatial structure of the SPOD modes. The mean component, for $k=0$, captures the block-to-block variation of the average velocity field and models deviations from the infinite-time averaged field. The spatial structures exhibit a distribution along the boundary of the main vortex inside the cavity and capture the aforementioned shear layer dynamics. In particular, the higher the frequency, the smaller the spatial length scale. 

\subsection{ROM configurations}\label{sec:ROM_configurations}
\begin{figure}
\centering
\includegraphics[width=\textwidth]{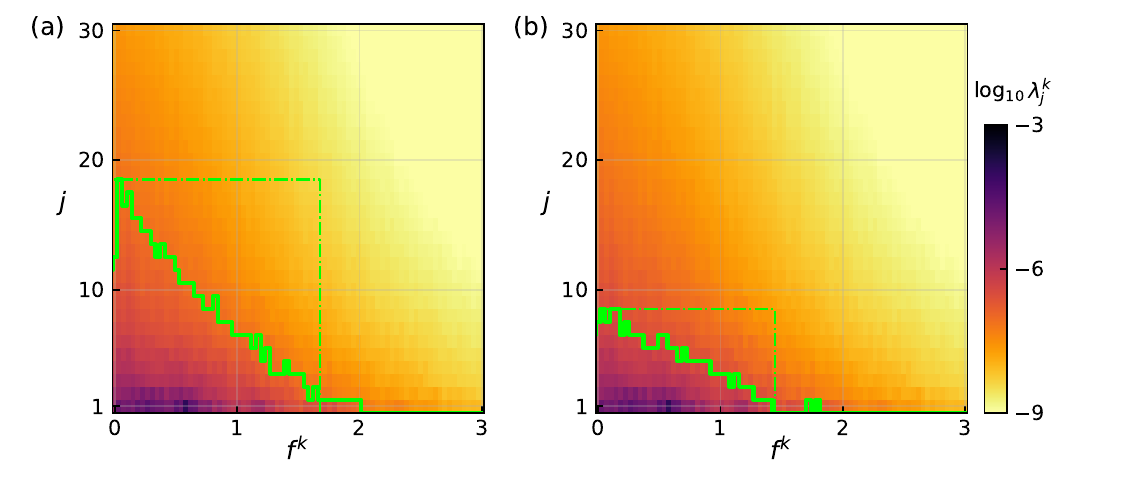}
\caption{Potential strategies for selecting SPOD modes for ROM R95, panel (a), and R90, panel (b), overlaid on the heat map of the SPOD eigenvalues. {The solid green boundary corresponds to the nominal truncation boundary,} where all eigenvalues are globally ranked prior to truncation. In practice, modes falling within the rectangular region enclosed by the green dash-dotted line are retained.}
\label{fig:ROM_truncation_in_SPOD_spectra}
\end{figure}

{The space-time nature of the proposed ROM framework implies that both frequency and modal truncation is needed. In principle, one could consider the entire frequency-mode spectrum, rank all SPOD modes by their eigenvalue and truncate low-energy modes regardless of their frequency or mode index to capture a predetermined fraction of the overall turbulent kinetic energy.}
\begin{table}
\begin{center}
\def~{\hphantom{0}}
\begin{tabular}{ccccccc}
ROM    &   $M$   &   $N$   & {No. SPOD modes} & Energy captured & Mode percentage & No. optimisation variables \\
[3pt]
R90 & ~8 &  36  & {~584} & 92.81\% & 2.95\% & ~585  \\
[3pt]
R95 & 18 &  42 & {1530} & 96.85\% & 7.73\% & 1531
\end{tabular}
\caption{Parameters for the two frequency domain ROMs used in the study.}
\label{tab:ROM_parameters}
\end{center}
\end{table}
{This strategy requires a varying number of SPOD modes at each frequency. For simplicity of computational implementation, we use the same number of modes at each frequency. The two panels of \cref{fig:ROM_truncation_in_SPOD_spectra} show the truncation boundary for two ROMs, referred to as R90 and R95 in what follows, designed to nominally capture 90\% and 95\% of the total energy. The heat maps show the base ten logarithm of the SPOD eigenvalues. The thick solid lines denote the truncation boundary of the mode indices and frequencies required to exactly recover $95\%$ and $90\%$ of the total energy using the ranking strategy, while the rectangular region bounded by the thin dashed-dotted lines defines the actual truncation boundary. Therefore, the actual overall energy contribution reconstructed here is slightly higher than the nominal value, as summarised in \cref{tab:ROM_parameters}.}

\section{Optimisation-based search of ROM solutions}\label{sec:ROM_optimisation}
\subsection{Constructing initial guesses}
\begin{figure}
\centering
\includegraphics[width=\textwidth]{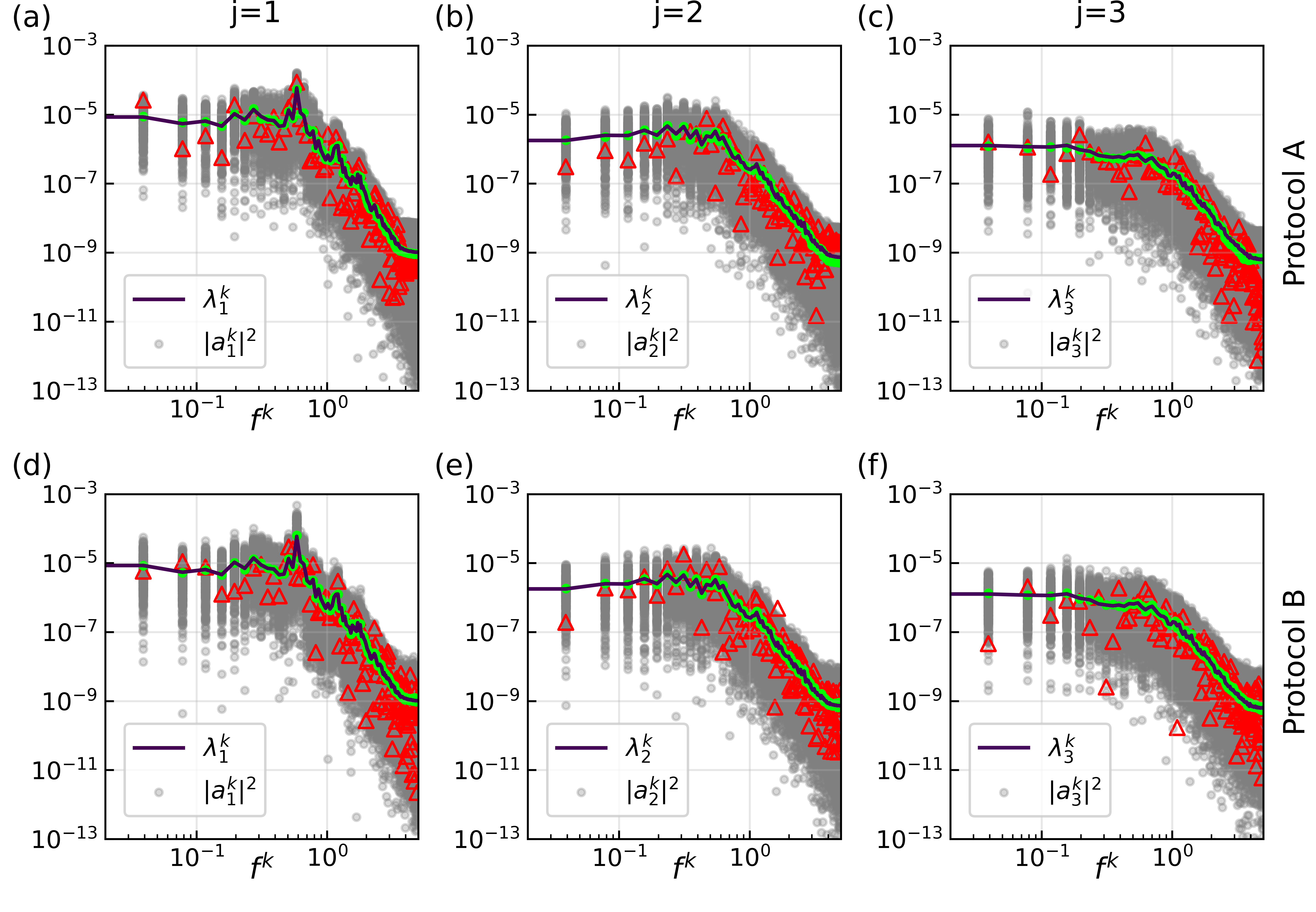}
\caption{The squared magnitude of the {300} sets of amplitude coefficients (light grey circles), obtained from protocol A (top panels) and protocol B (bottom panels). These are compared to the SPOD eigenvalues (solid lines) for the first three SPOD modes. The red triangles denote coefficients from one of the initial guesses and the green dots denote the average over all guesses. }
\label{fig:amplitude_coefficients_squares}
\end{figure}

Providing appropriate initial guesses for the search is crucial to finding physically relevant ROM solutions in the low-order subspace. Conventional methods for finding UPOs are usually initialised with DNS states obtained by recurrence analysis~\citep{Chandler_Kerswell_2013,page_recurrent_2024} because turbulent trajectories are believed to visit invariant solutions in state space~\citep{crowley_turbulence_2022}.
It may be argued that, in the present case, segments of DNS trajectories that produce near-recurrence events may be situated, when projected onto the low-order subspace, close to the projection of a neighbouring UPO on the same subspace. Hence, in principle, given that a fixed period $T = 25.6$ is used for SPOD analysis, one option may be to find near-recurrences with such a period from the available DNS data. However, this approach would only generate a limited number of initial guesses. In practice, two alternative approaches were considered. In the first approach, referred to as protocol A, we {partition the same DNS data used for SPOD into $300$ blocks of data of length of $T=25.6$, using an overlap of $87.5\%$ and then project their Fourier transform} onto the space-time basis functions by computing
\begin{equation}
a_j^k = \left(  \bm{\phi}_{j}^{k}, \bm{\hat{u}}'_k \right)  
\, ,
\end{equation}
where $\bm{\hat{u}}'_k$ denotes the Fourier coefficients {of the transform} of $\bm{u}'$ at frequency $f^k$, leading to {$300$} sets of projected amplitude coefficients that are used as initial guesses. Clearly, not all of these blocks of DNS data correspond to near-recurrence events. The amplitude coefficients of the first, second, and third SPOD modes obtained with this method are shown in the top panels of \cref{fig:amplitude_coefficients_squares}, compared to the spectrum of the SPOD eigenvalues. The projected coefficients are scattered around the SPOD eigenvalues, as their average at each frequency and mode pair must be equal to the corresponding eigenvalue, according to the properties of the SPOD method. However, significant differences in amplitude (and phase) are observed between sets of amplitude coefficients, providing a rich set of initial guesses.
The second approach is motivated by the fact that gradient-based optimisers, employed here to solve the optimisation problem, may be quite sensitive to initial guesses. To investigate the robustness of the proposed method, we also introduce protocol B, which consists of generating random amplitude coefficients (with random amplitude and phase) sampled from Gaussian distributions with mean and standard deviation derived from the projected amplitude coefficients from protocol A, for each frequency-mode pair. {This approach is analogous to the method recently discussed in \citet{beck} to generate initial guesses for the search of UPOs, i.e.~by producing via an order reduction method time-periodic space-time fields that lie on the chaotic attractor and match the statistics of the system.} For consistency, we generate {300} initial guesses using this second approach. The bottom panels of \cref{fig:amplitude_coefficients_squares} show the distribution of the coefficients for the same SPOD modes generated using this protocol. Despite the similarity between the spectra of initial guesses generated using the two protocols, it will be shown later that the solutions obtained via the optimisation initialised from these two sets of guesses have important differences.

\subsection{Optimisation results}
\begin{figure}
\centering
\includegraphics[width=\textwidth]{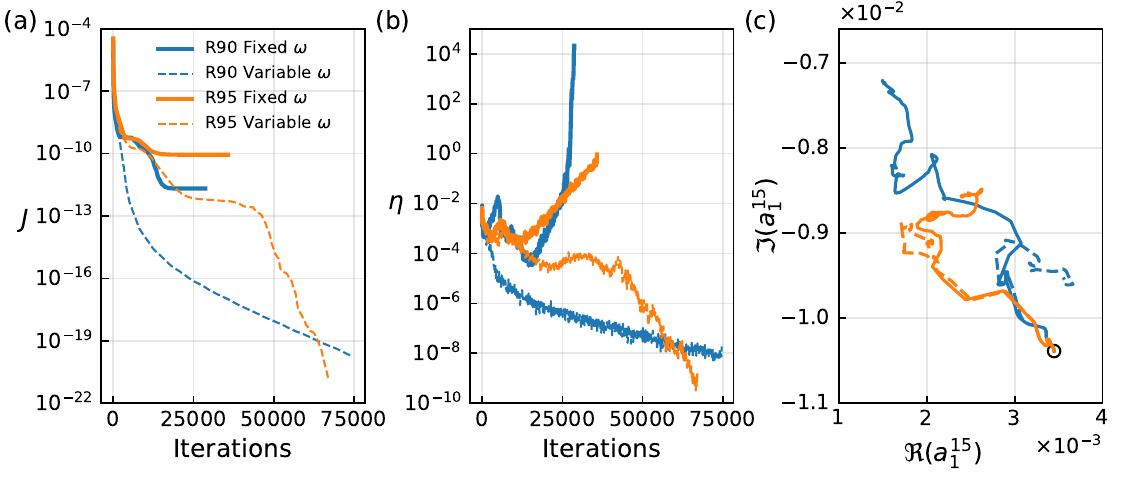}
\caption{Optimisation history of the objective function $J$, panel (a), and the metric $\eta$ for the ROM R90 and R95 using initialisation of protocol A projected from one data block, with and without optimising time periods $\omega$, panel (b). Optimisation history of the dominant amplitude coefficient at $f^{15}=0.587$, with the initial guess marked with an empty circle, panel (c).}
\label{fig:Obj_Metric_comp}
\end{figure}

The convergence of the optimisation problem for the two ROMs is first demonstrated using an initial guess from protocol A, with and without inclusion of the fundamental frequency $\omega$ as an optimisation variable. \cref{fig:Obj_Metric_comp}(a) shows the history of the objective function $J$. In the early stages of the optimisation, in the first one thousand iterations, the convergence history for both ROMs shows a similar trend regardless of whether $\omega$ is included in the set of optimisation variables. Here, the objective function and its gradient are reduced at a similar rate and the metric $\eta$ remains at a similar level of magnitude, as shown in \cref{fig:Obj_Metric_comp}(b). {Note that $\eta$ appears to be stochastic because, while the objective function is reduced monotonically between iterations, the gradient norm does not obey the same property.}
However, optimisation over $\omega$ becomes more crucial as optimisation progresses. When $\omega$ is fixed, the optimiser converges to a {non-zero} minimum (i.e.~not a solution of the ROM, although the residual has decreased substantially before convergence) and the metric $\eta$ diverges in the final iterations. In contrast, changing $\omega$ enables a further reduction of the objective function to arbitrarily low values, for both ROMs. In this case, the metric $\eta$ decreases significantly, indicating that this solution indeed satisfies the low-order system.  \cref{fig:Obj_Metric_comp}(c) compares the optimisation history of the amplitude coefficient at the dominant peak frequency of $f^{15}=0.586$ for the four scenarios examined here. The optimisation follows different routes, implying that the final optimal solutions are substantially different when the fundamental frequency is included in the optimisation. This phenomenon was also observed for all other available initial guesses, all of which converged to {non-zero} minima when fixing the fundamental frequency. This result is akin to the fact that the period of UPOs of the Navier-Stokes equations, but also for chaotic systems in general, is not known a priori, but must be found during the search process. Hence, we always included the fundamental frequency $\omega$ in subsequent studies.

\begin{figure}
\centering
\includegraphics[width=\textwidth]{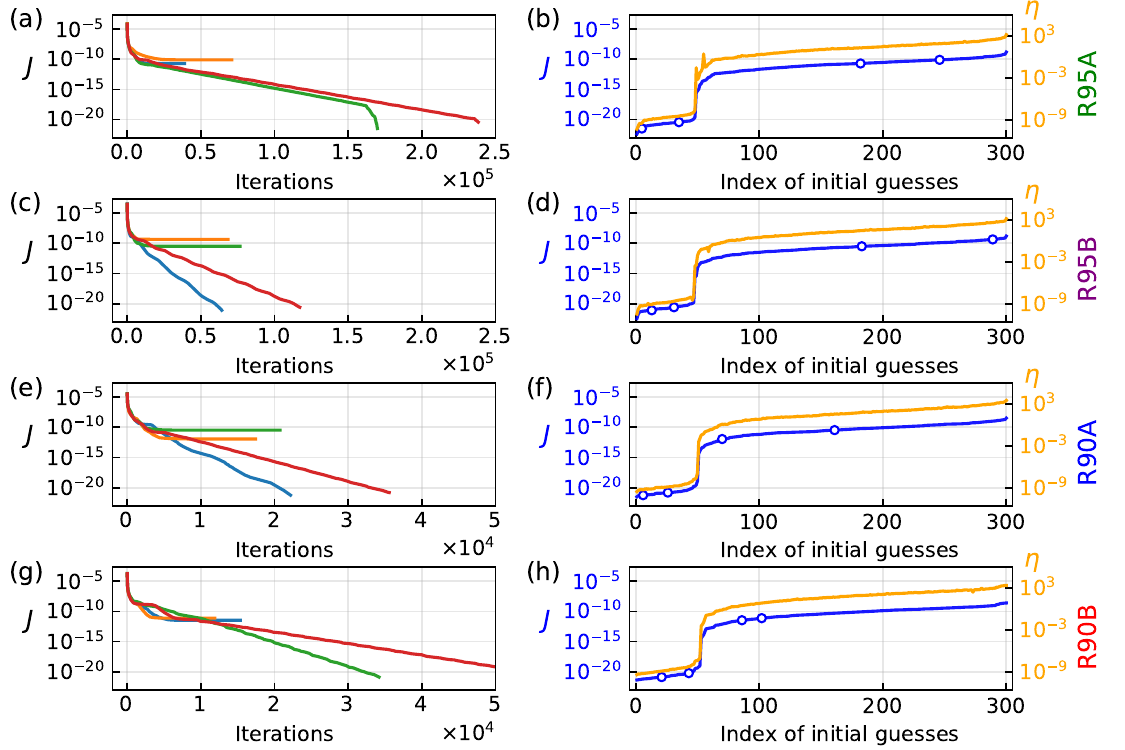}
\caption{Optimisation history of the objective function $J$ with variable $\omega$ for {the two ROMs and for a few initial guesses from protocols A and B, left panels}. The final objective function values, sorted in ascending order, and the metric $\eta$ are plotted in the right panels.}
\label{fig:Obj_all_block}
\end{figure}

{The left panels of \cref{fig:Obj_all_block}} show the convergence history for the two ROMs {for a random subset of} the initial guesses available from the two protocols. Part of the initial guesses lead to actual ROM solutions, while others lead to {non-zero} minima even if $\omega$ is allowed to vary. The value of the objective function {$J$} and the metric $\eta$ at the point of convergence, sorted in ascending order for all {the 300 initial guesses}, are shown in the right panels of \cref{fig:Obj_all_block}. This criterion may be employed to distinguish actual solutions from {non-zero} minima, although it may fail for solutions that exhibit a much slower than average convergence rate.
{Overall, multiple ROM solutions are obtained from solving the optimisation problem in the four cases, as summarised in \cref{tab:APOs_J_0}. The success rate of the search does not seem to vary significantly between the two protocols, nor between the two ROMs. Specifically, about one initial guess in six leads to a ROM solution, as the majority of guesses converge to  {non-zero} minima, analogous to the ghost states recently proposed by \citet{zheng2024ghost}.}



\begin{table}
\begin{center}
\def~{\hphantom{0}}
\begin{tabular}{cccc}
ROM & Initialisation & Number of solutions & Success rate  \\[3pt]
\multirow{2}{*}{R90} & Protocol A & {48} & {16.0\%}  \\
                     & Protocol B & {47} & {15.7\%}  \\ [3pt]
\multirow{2}{*}{R95} & Protocol A & {50} & {16.7\%}  \\
                     & Protocol B & {51} & {17.0\%}
\end{tabular}
\caption{Success rate of finding the ROM's solutions by solving the optimisation problem with a variable $\omega$ for ROM R90 and R95, using initial guesses of protocol A and B.}
\label{tab:APOs_J_0}
\end{center}
\end{table}

\subsection{Uniqueness of ROM solutions}\label{sec:APO_uniqueness}
To quantify whether the ROM solutions are unique and to remove duplicates, a ``hash" function is defined that assigns a single real number to each solution. Duplicates can then be found when the hash function evaluated on two solutions produces the same output, up to a small tolerance. Several options may be devised. For example, the frequency $\omega$ would be a good choice as the periods of UPOs in a chaotic system are, up to symmetries, all distinct. Here, we opted for the energy-like quantity
\begin{equation}
\begin{aligned}
\xi &=  \sum_{j=1}^{M} \sum_{k=-N}^{N} | a_j^k|^2
\, .
\end{aligned}
\label{equ:unique_value_ROM}
\end{equation}
Using this definition, rather than directly comparing the amplitude coefficients, avoids the potential issue arising when two identical solutions differ only by a phase shift.

\begin{figure}
\centering
\includegraphics[width=\textwidth]{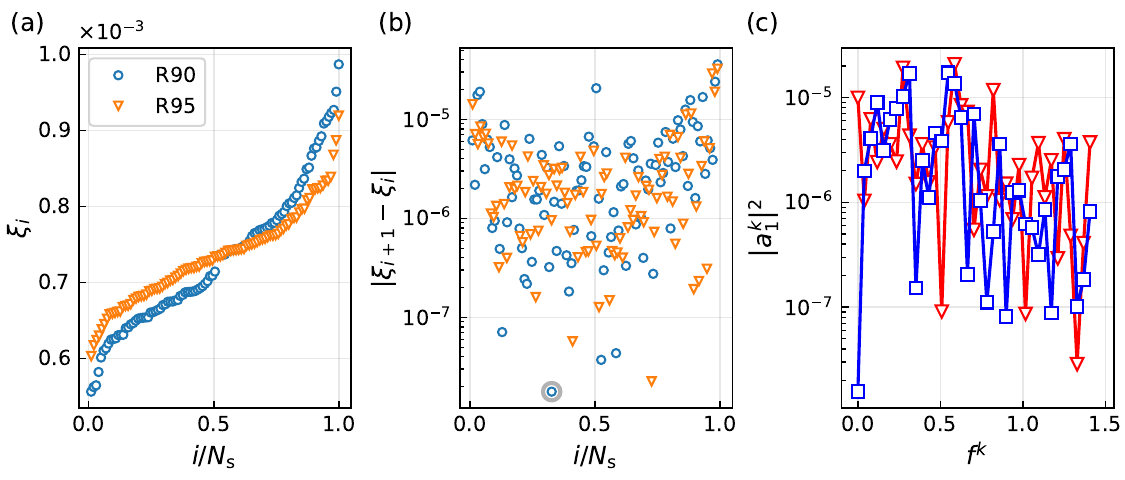}
\caption{The hash function evaluated on the solutions obtained for the two ROMs, in ascending order, panel (a). {The results combine the solutions obtained using initial guesses from the two protocols}. Difference between the hash function evaluated on the solutions, panel (b). Squared magnitude of the amplitude coefficient $a_1^k$, for the two ROM solutions with the lowest difference in the hash function, panel (c), as marked by the grey circle in panel (b).}
\label{fig:unique_value_90ROM_95ROM}
\end{figure}

\cref{fig:unique_value_90ROM_95ROM}(a) shows the values of $\xi$ for all ROM solutions obtained from the ROM R90 and R95, sorted in ascending order{, $N_s$ denotes the number of zero-minimum solutions found. The results include} solutions obtained using initial guesses from protocol A and protocol B to identify potential duplicates that might arise from different ways of initialisation.
\cref{fig:unique_value_90ROM_95ROM} (b) shows the difference of $\xi$ between adjacent pairs of solutions, where the smallest difference is on the order of $10^{-8}$. This is still relatively high compared to the convergence tolerance with which solutions and their hash functions are obtained in the optimisation. In fact, we observed that further reducing the convergence tolerance for the optimisation did not result in significant changes to the hash function of the solutions or their differences. To further demonstrate this aspect, \cref{fig:unique_value_90ROM_95ROM}(c) compares the squared magnitude of the amplitude coefficients of the first SPOD modes as a function of the frequency for the two solutions with the lowest difference of the hash function, indicated in the panel (b) by a grey circle. A significant discrepancy is observed at different discrete frequencies, indicating that these solutions are in fact distinct, although the difference in the hash function may appear small. Therefore, we conclude that all solutions obtained for ROM R90 and R95 are unique. This was initially unexpected because it is not uncommon to find duplicate UPOs for chaotic systems. However, this may be explained by the fact that the period of solutions sought is quite high ($T=25.6$), and is several times longer than the period of dominant oscillation in the cavity ($T_{\mathrm{osc}} \approx 1.7$). In the present optimisation-based context, the number of ROM solutions is related to the non-convexity of the objective function $J$ of equation \eqref{equ:NS_incomp_SPOD_proj_model_res_obj}. Inspecting \eqref{equ:NS_incomp_SPOD_proj_model} suggests that the frequency $\omega$ plays an important role in determining the relevant strength of the linear and quadratic terms in the ROM, and thus of the quadratic and quartic terms in the objective function. When $\omega$ is large -- corresponding to short time periods -- the linear term is dominant, and thus the objective function has a strong quadratic behaviour. Conversely, when $\omega$ is small -- corresponding to long periods -- the quadratic terms become more important and the objective function may display strong quartic behaviour, admitting a larger number of minima and thus a larger number of solutions. This perspective is consistent with evidence that the number of UPOs in chaotic systems increases exponentially with the period~\citep{davidchack_efficient_1999}.

\begin{figure}
	\centering
	\includegraphics[width=\textwidth]{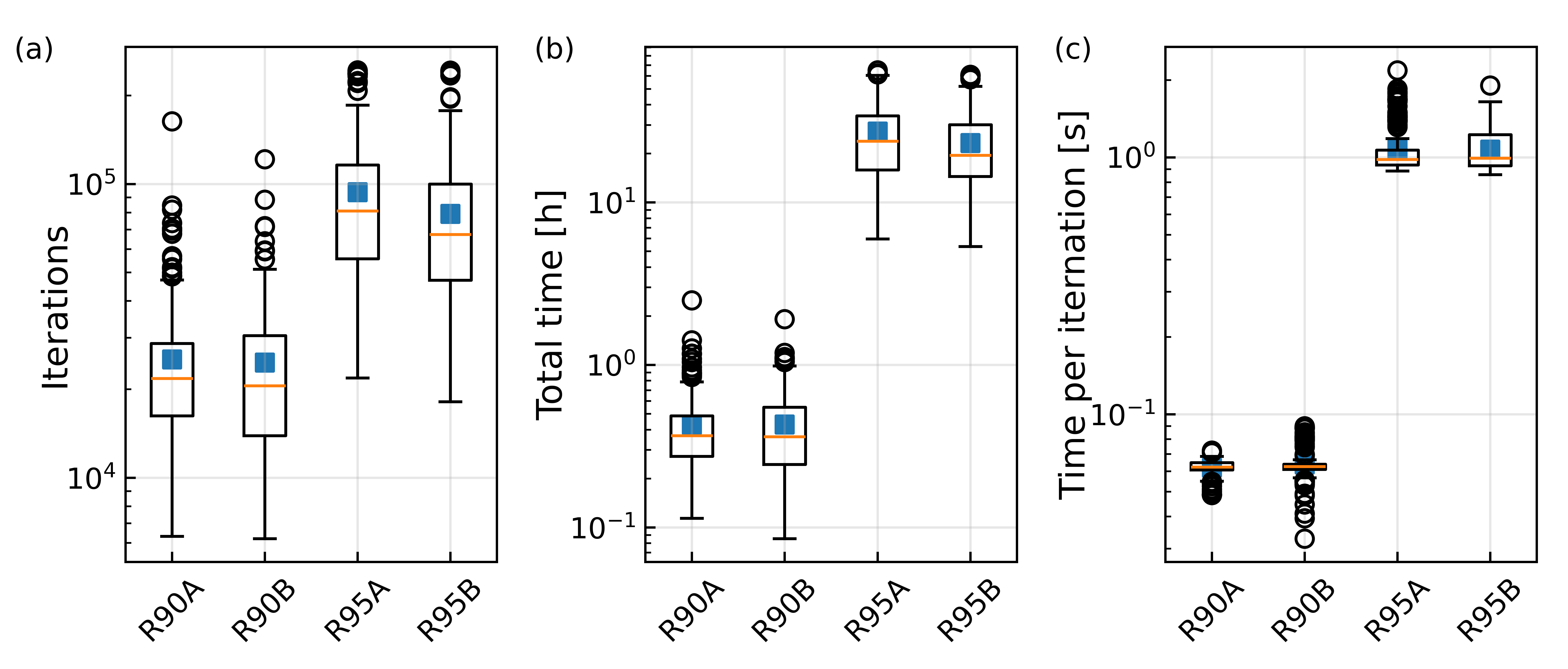}
    \caption{{Box-and-whisker plots of (a) total number of iterations, (b) total computational time and (c) computational cost of each iteration for the ROM R90 and R95 using initial guesses of protocol A and B, respectively. The boundaries of the whiskers are based on the distance of 1.5 times the inter-quartile range. The blue squares denote the average for each case.}}
	\label{fig:ROM_computing_time}
\end{figure}

{
\subsection{Computational cost of finding ROM solutions}\label{sec:computational_time}
The computational cost associated to finding the ROM solutions depends primarily on the evaluation of the convolutions in the nonlinear term expressed by the model coefficient tensor $\bm{Q}$ in \cref{equ:NS_incomp_SPOD_proj_model}. This cost scales as $\mathcal{O}(M^3 N^2)$, sharing the cubic dependence on the number of modes with space-only ROMs, but featuring a quadratic scaling in time. In practice, the nonlinear term needs to be evaluated whenever the optimiser computes the violation of the governing equations and its gradient with respect to the amplitude coefficients. \cref{fig:ROM_computing_time} shows a summary of the costs associated with finding ROM solutions on a computer with a dual socket AMD EPYC 9654 processor, showing the total number of iterations, the total computational cost and the computational time of each iteration in box-and-whisker plots. Both the total number of iterations and the computational time per iteration increase when the ROM includes more optimisation variables, from $585$ in R90 to $1531$ in R95. The average time per iteration is similar for the optimisation using initial guesses of protocol A and B in R90 or R95. The ratio of the averaged iteration time between the ROM R95 and R90 is $15.86$ for protocol A and $16.81$ for protocol B, which agrees well with the expected $\mathcal{O}(M^3 N^2)$ scaling, i.e. $15.50$. The actual computational cost also depends on the design and implementation of the program structure, which uses serial computation in the present case. Techniques to alleviate these costs, e.g.~evaluating the nonlinear term in physical space, exploiting sparsity or using Discrete Empirical Interpolation Method (DEIM), have been discussed in the literature \citep{frame_space-time_2024}, but have not been considered in the current work. Additionally, using the Newton-Raphson method, rather than an optimisation-based approach, should accelerate convergence significantly, further reducing costs, as gradient-based methods are known to have slow convergence rates \citep{azimi_constructing_2022, Burton2025ResolventBased}.}

\section{Analysis of ROM's solutions}\label{sec:APO_results}
We now investigate the nature of the ROM solutions and evaluate their ability to model the dynamics and statistics of the original system.

\subsection{Energy spectrum}\label{sec:APO_spectra}
The ensemble-averaged energy spectrum of all solutions is computed for each frequency-mode pair as
\begin{equation}
\begin{aligned}
E_j^k &= \frac{1}{N_\mathrm{s}} \sum_{n=1}^{N_{\mathrm{s}}}\left| a_{j}^{k}(n) \right|^2
\, ,
\end{aligned}
\label{equ:ROM_averaged_spectrum}
\end{equation}
where $a_j^k(n)$ denotes the amplitude coefficients of the $n$-th solution and $N_{\mathrm{s}}$ is the total number of solutions found. Note that this averaging procedure neglects the fact that the fundamental frequency $\omega$ differs {slightly} between solutions. \cref{tab:Perioid_ROM_solutions} summarises basic statistics of this quantity for the four scenarios investigated, including data for initial guesses that converged to a {non-zero} minimum, for completeness. It is found that the maximum deviation {of ROM solutions} from the reference value obtained from the SPOD analysis ($\omega = 2\pi/25.6  \simeq  0.24544$) is less than {2.3\%}, but most solutions deviate much less.
\begin{table}
\begin{center}
\def~{\hphantom{0}}
\begin{tabular}{lllccc}
                       & Solution type &  Mean   & Standard deviation & Minimum & Maximum \\[3pt]
\multirow{2}{*}{R95A} & ROM solutions & 0.24565 & 0.00169 & 0.24107 & 0.24982  \\
                       & {Non-zero} minima  & 0.24549 & 0.00163 & 0.23992 & 0.24926  \\
\multirow{2}{*}{R95B} & ROM solutions &  0.24580 & 0.00180 & 0.24137 & 0.24952  \\
                       & {Non-zero} minima  & 0.24585 & 0.00166 & 0.24107 & 0.25110   \\
\multirow{2}{*}{R90A} & ROM solutions & 0.24405 & 0.00198 & 0.23963 & 0.24871  \\
                       & {Non-zero} minima  & 0.24396 & 0.00171 & 0.23858 & 0.24882  \\
\multirow{2}{*}{R90B} & ROM solutions &  0.24397 & 0.00189 & 0.23835 & 0.24814  \\
                       & {Non-zero} minima  &0.24418 & 0.00178 & 0.23791 & 0.25000 \\
\end{tabular}
\caption{The mean, standard deviation, minimum and maximum of the fundamental frequencies obtained from the ROM R95 using the initial guesses of protocol A (R95A), {R95 using the initial guesses of protocol B (R95B)}, the ROM R90 using the initial guesses of protocol A (R90A), and the ROM R90 using the random initial guesses of protocol B (R90B). Statistics are shown for the ROM's solutions ($J=0$) and {non-zero} minima.}
\label{tab:Perioid_ROM_solutions}
\end{center}
\end{table}
Consequently, the variation of each discrete frequency resolved by the ROM also remains within this range, and such differences are neglected when computing the ensemble spectrum.

\begin{figure}
\centering
\includegraphics[width=\textwidth]{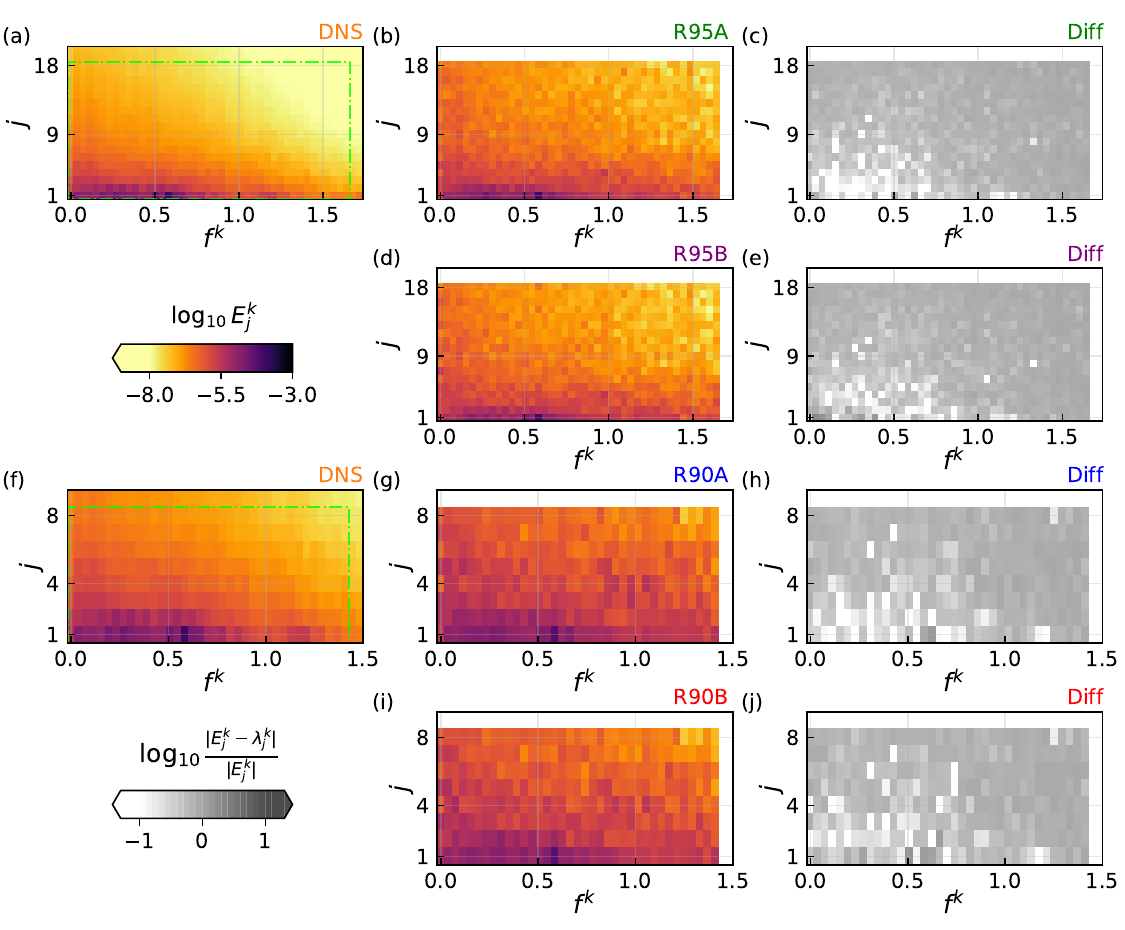}
\caption{Comparison of the ensemble-averaged spectrum $ E_j^k $ between the SPOD eigenvalue spectrum, panels (a, {f}) and the solutions obtained from the ROM R95 with protocol A, panel (b), {R95 with protocol B, panel (d),} R90 with protocol A, panel ({g}), and R90 with protocol B, panel ({i}). { The relative error of the ensemble-averaged spectrum is shown in the panels in the last column. The dash-dotted rectangles in the left panels indicate the truncation boundary of the two ROMs.}}
\label{fig:spectral_90ROM_95ROM}
\end{figure}

Figures \ref{fig:spectral_90ROM_95ROM}({b, d, g, i}) show the ensemble-averaged spectrum of the ROM solutions compared with the SPOD eigenvalue spectrum in panels (a) and {(f)}.
The ROM solutions are observed to capture relatively well the frequency of the dominant peak and the energy distribution of low-index modes ($j \le 6$ for R90 and $j \le 10$ for R95) at low and medium frequencies ($f^k \le 0.7$). In contrast, the spectral energy at high frequencies ($ f^k>0.7$) and high-index SPOD modes ($j>6$ for R90 and $j>10$) for R95 is overestimated {as shown in the right panels in \cref{fig:spectral_90ROM_95ROM}}, although ROM R95 is slightly more accurate in predicting the energy content in these regions.
{The large relative error is due to the low energy at high frequencies.}
It may be argued that this behaviour is due to the truncation. Studies focused on classical POD-Galerkin space-only models (see \citet{holmes_turbulence_2012,balajewicz_low-dimensional_2013,chua_khoo_sparse_2022}) suggest that the truncation of small spatial scales results in the accumulation of energy in high-index modes. In the present case, truncation is applied across both the spatial and temporal directions, resulting in a greater set of nonlinear energy interactions between resolved spatio-temporal scales and the unresolved scales discarded by the model. The higher energy near the truncation boundaries may therefore represent a compensatory mechanism to maintain the energy balance. However, it is unclear how truncation along the modal or temporal direction may result in higher energy throughout the spectrum, as minimisation of the residual norm, the objective function \eqref{equ:NS_incomp_SPOD_proj_model_res_obj}, couples all length and temporal scales together. In any case, the spectral energy at the most dominant frequency-mode pairs is captured with a satisfactory degree of accuracy.

\begin{figure}
\centering
\includegraphics[width=\textwidth]{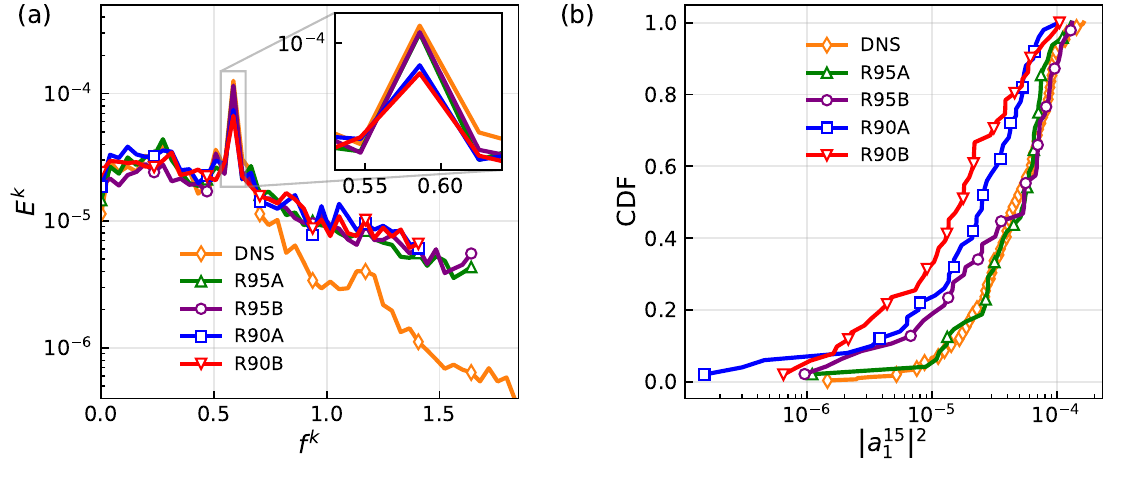}
\caption{Ensemble-averaged spectrum for solutions obtained from the ROM R95 with protocol A~(R95A), {R95 with protocol B~(R95B),} R90 with protocol A~(R90A) and R90 with protocol B~(R90B), compared to the DNS data, panel (a). Cumulative distribution function (CDF) of the squared magnitude of the amplitude coefficient $a_1^{15}$ for different ROM solutions compared to DNS, panel (b){, with data shown as symbols every 5 points}.}
\label{fig:spectral_90ROM_95ROM_freq}
\end{figure}

A further reduction is performed by summing the energies of all modes at the same frequency, i.e., by computing
\begin{equation}
\begin{aligned}
E^k & = \sum_{j=1}^{M} E_j^k
\; .
\end{aligned}
\end{equation}
This quantity is reported in \cref{fig:spectral_90ROM_95ROM_freq}(a) for the {four} ROM scenarios investigated. 
{
In all cases, the spectral energy for frequencies below $0.5$ is well captured, but at frequencies above 0.7, after the peak, the ROM solutions over-predict the energy levels and do not show a peak at the second harmonic frequency, which is probably too weak to be resolved by the model. Despite this, the frequency $f^{15}=0.587$ of the dominant peak is captured correctly, though with some differences in its energy across the models.
To better illustrate this fact, the Cumulative Distribution Function (CDF) of the squared magnitude of the amplitude coefficient at the dominant peak frequency ($a_1^{15}$) is compared with that obtained from DNS for the 300 data blocks in \cref{fig:spectral_90ROM_95ROM_freq}(b). For ROM R95, it is found that only solutions obtained from protocol A model well the DNS distribution, while solutions from protocol B are more likely to predict a lower energy at the dominant frequency. On the other hand, solutions of ROM R90 have consistently lower energy at the dominant frequency, with solutions from protocol B performing worse. The difference between the two protocols suggests that the objective function \eqref{equ:NS_incomp_SPOD_proj_model_res_obj} may indeed have a very large number of zero-minima but some of those found from protocol B may be non-physical or may fail to capture the correct amplitude of the dominant flow patterns. Nevertheless, it should be noted that these results are achieved without resorting to model calibration or closure techniques \citep{ahmed_closures_2021}. These are typically needed for space-only POD-Galerkin models, which otherwise tend to significantly overpredict the energy spectral content, with some exceptions \citep{Cavalieri}.}



\subsection{Statistics and dynamics of ROM solutions}\label{sec:APO_statistics}
\begin{figure}
\centering
\includegraphics[width=\textwidth]{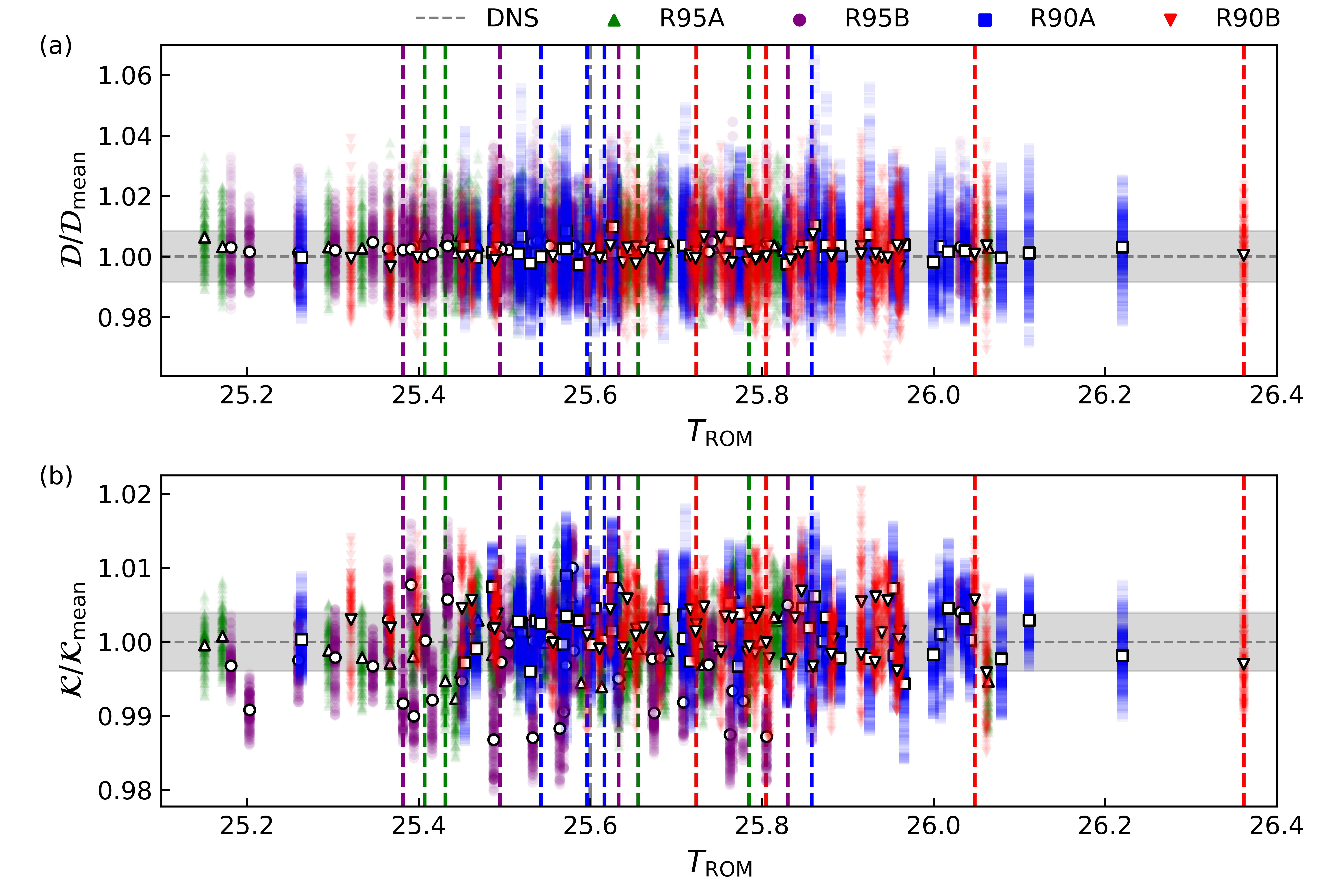}
\caption{The distribution of energy dissipation rate $\mathcal{D}$, panel (a), and kinetic energy $\mathcal{K}$, panel (b), for the solutions obtained in the {four} ROM cases. For each solution, the period average is shown with a white symbol. The horizontal dashed grey line and the grey region represent the mean value and standard deviation from DNS.}
\label{fig:phase_space}
\end{figure}

\begin{figure}
\centering
\includegraphics[width=\textwidth]{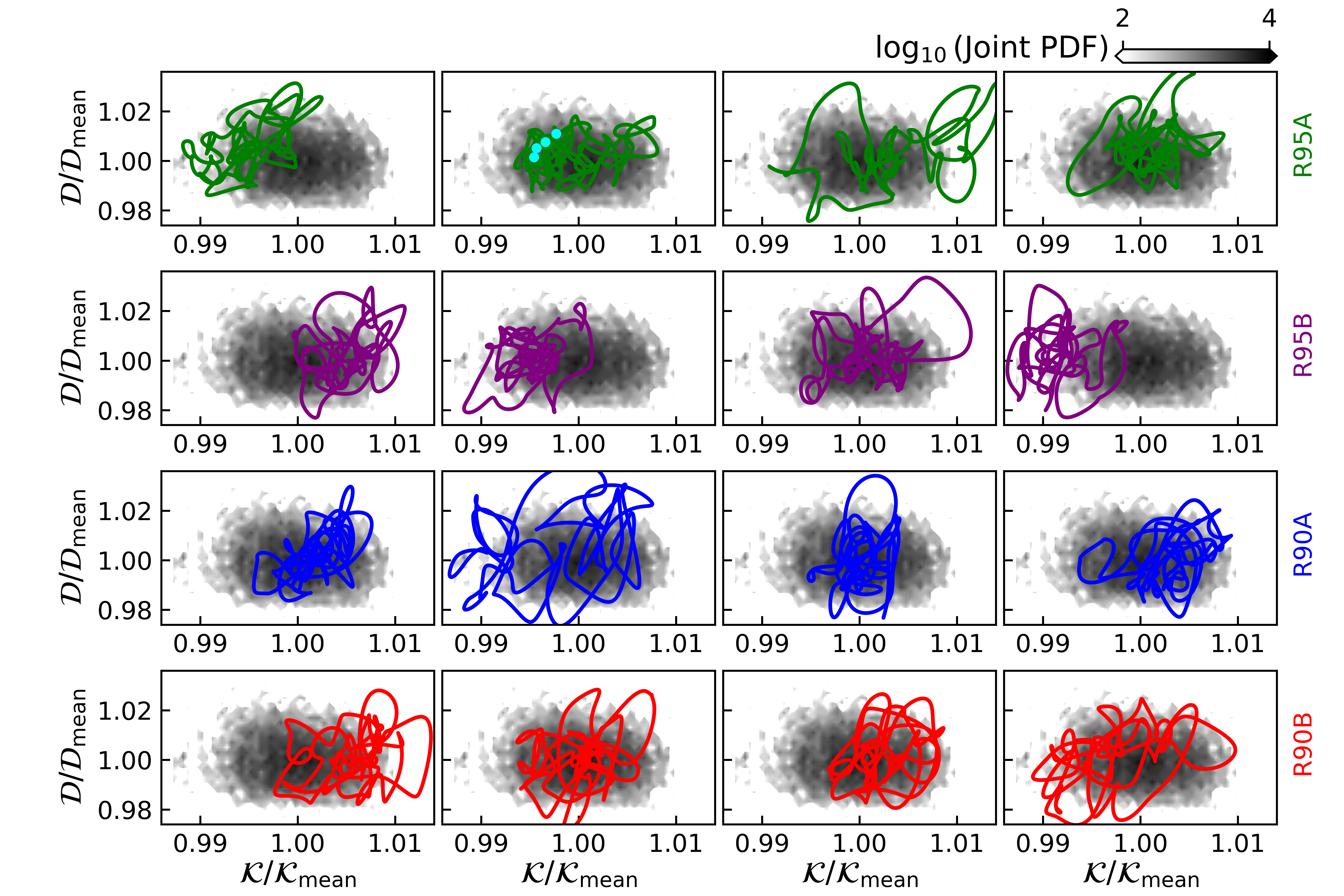}
\caption{Projection on the kinetic energy and dissipation rate plane for four selected trajectories in each ROM, indicated by dashed lines in \cref{fig:phase_space}, superimposed to the joint PDF from DNS run over $1000$ time units. Both $\mathcal{K}$ and $\mathcal{D}$ are normalised by the mean DNS value. {Four solid cyan dots} represent instantaneous states sampled to display the associated vorticity fields in \cref{fig:vortex_optimal}.}
\label{fig:phase_space_traj}
\end{figure}

\cref{fig:phase_space}(a) shows data for the energy dissipation rate, $\mathcal{D} = \| \nabla \times \bm{u} \|^2 / Re$, obtained from the reconstruction of the velocity field derived from the available ROM solutions, plotted with respect to the time period. For each solution, data are sampled evenly over the period and the period average dissipation rate is reported using a white symbol. The data is normalised by the mean quantity computed from a long DNS run for $1000$ time units. The horizontal grey region represents the standard deviation of the DNS data. The same data is shown for the kinetic energy in panel (b). We observe that, in most cases, the period average computed from the ROM solutions falls in a range that is one standard deviation wide around the mean value. There is no discernible effect of the period on these statistics, as often observed for short UPOs of chaotic systems \citep{lasagna_sensitivity_2020}. Here, all solutions are relatively long compared to the period of the shear layer dominant oscillation. {The trajectories of a few selected solutions, indicated by the dashed vertical lines in panel \cref{fig:phase_space}(a), and projected onto the $\mathcal{K}-\mathcal{D}$ plane are shown in \cref{fig:phase_space_traj} along with the joint probability density function of these quantities computed from DNS data (shown as a grey heat map). The solutions shown in the figure are representative of several classes of solutions -- of differing abundance -- which are distinct both in their location on the $\mathcal{K}-\mathcal{D}$ plane and in their variance. The solutions found appear to span nearly the full range of states visited by DNS, as far as this simple two-dimensional projection can show, and frequently visit the high-probability states explored by DNS. Examination of the complete set of these projections shows that solutions obtained from protocol A fall in the high-probability region of the PDF from DNS with much greater likelihood than those from protocol B, especially for ROM R95. For R95B, some solutions appear to reside outside of the attractor of the full-order system, even though this effect is moderate. This is the case, for instance, of the first and second examples for R95B in the figure, where lower or higher than average kinetic energy is observed. Yet, as discussed in \cref{fig:phase_space}, the period average of flow quantities for most ROM solutions is close to the DNS data. This suggests that the proposed ROM can indeed predict the amplitude of dominant dynamical features with reasonable accuracy, despite no closure model being utilised to correct the long-term behaviour of the model.} 

Four snapshots of the vorticity fields sampled at $\Delta t/T_{\mathrm{ROM}}=1/64$ in the second trajectory for R95A, marked by the solid dots in \cref{fig:phase_space_traj}, are shown in \cref{fig:vortex_optimal}. The animations for the entire period are presented in the supplementary material. The main vortex and the shear layer are well captured by this ROM solution, compared to the vorticity field of DNS in \cref{fig:vorticity_timeSeries}(a). The ROM solution is capable of depicting the strong interaction between the shear layer and the corner flows, specifically the erratic roll-up of vorticity and the shedding of structures from the corners. However, it can also be noticed that small-scale structures have an excessive amplitude in the velocity fields obtained from the reconstruction. We argue that this is the manifestation of the overestimation of the energy of the high-frequency and high-index modes, as previously discussed for the spectra in \cref{fig:spectral_90ROM_95ROM}. 
\begin{figure}
\centering
\includegraphics[width=\textwidth]{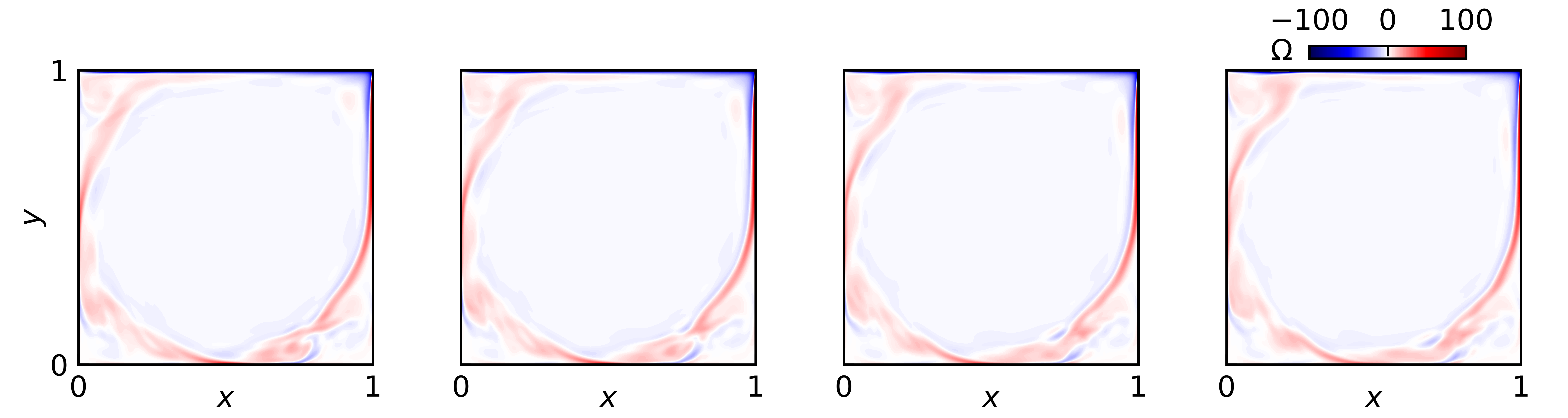}
\caption{Snapshots of the vorticity field sampled at $\Delta t/T_{\mathrm{ROM}}=1/64$ {from the second solution of ROM R95 obtained using the initial guesses of protocol A, corresponding to the states marked by solid dots in the corresponding panel of \cref{fig:phase_space_traj}.}}
\label{fig:vortex_optimal}
\end{figure}

\begin{figure}
\centering
\includegraphics[width=\textwidth]{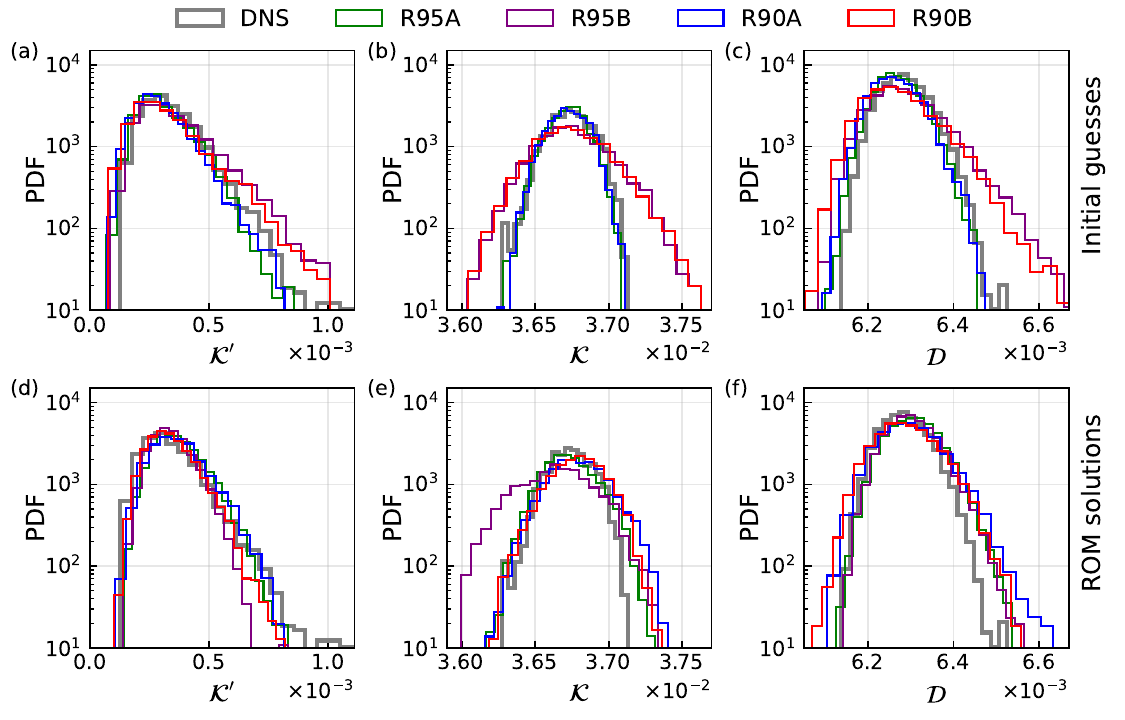}
\caption{{The probability density function (PDF) of turbulent kinetic energy $\mathcal{K}'$, kinetic energy $\mathcal{K}$ and energy dissipation rate $\mathcal{D}$ obtained from the reconstruction of the velocity field using the initial guesses from the two protocols (top panels) and from the ROM solutions (bottom panels), compared with distributions from DNS data.}}
\label{fig:PDF_KE_TKE_D_I}
\end{figure}

{To get a more detailed view on the statistics of flow quantities predicted by the ROM, we compare in \cref{fig:PDF_KE_TKE_D_I} the PDFs of the turbulent kinetic energy $\mathcal{K}'$, kinetic energy $\mathcal{K}$ and energy dissipation rate $\mathcal{D}$ with those obtained from DNS. PDFs are constructed by sampling the time-periodic velocity field reconstructed from each set of amplitude coefficients with a temporal resolution of $\Delta t/T_{\mathrm{ROM}}=1/256$ and then aggregating the samples of all solutions. The results of the reconstructions using the ROM solutions are shown in the bottom panels, while, for completeness, the PDFs of the reconstructions from the initial guesses obtained directly from projection in protocol A or from the random guesses in protocol B are shown in the top panels. The initial guesses from protocol B, derived from the statistics of the projection coefficients, produce probability distributions with noticeably wider support, highlighting a limitation of the random-generation approach. On the other hand, the PDFs from protocol A are, as expected, close to the long-time DNS statistics, apart from modest differences attributable to the modal truncation. We note that none of these guesses satisfies the reduced order model. After the optimisation, the PDFs of the reconstructions with the ROM solutions match the DNS statistics relatively well. An exception is ROM R95 using guesses from protocol B, which shows a broader left tail in the distribution of the kinetic energy, as several solutions display lower-than-average kinetic energy (as the fourth example in \cref{fig:phase_space_traj}). ROM R90 with guesses from protocol B, or using guesses from protocol A, does not exhibit the same behaviour, which could be due to the fact that the higher-dimensional model has a larger number of zero-minima corresponding to non-physical solutions, more easily found with guesses from protocol B. Modest differences in the tails of the distributions are observed, but overall the ROM solutions reproduce well the distributions of flow statistics.}

\section{Conclusions}\label{sec:conclusions}

We have developed a space-time nonlinear Reduced-Order Modelling~(ROM) framework specifically designed for statistical stationary fluid flows, extending the space-time ROM approach proposed in earlier work~\citep{choi_space--time_2019,frame_space-time_2024}. Unlike conventional space-only ROMs, the method employs space-time basis functions that represent dominant spatio-temporal coherent structures to achieve order reduction in both space and time. Although several choices may exist, Spectral Proper Orthogonal Decomposition~(SPOD) was used here to generate the reduced basis. Projection of the Navier--Stokes equations onto the modes using a space-time inner product yields a ROM in the form of a nonlinear algebraic system governing the amplitude coefficients of the basis functions. To solve the system, we proposed to use a robust gradient-based optimisation strategy, inspired by recent adjoint-based methods for locating invariant solutions of the Navier-Stokes equations, that enforces conservation laws within the reduced subspace.

Numerical experiments on two-dimensional lid-driven cavity flow at $Re=20{,}000$, a regime characterised by chaotic dynamics, are used to investigate the performance of the proposed method. Long-period ROM solutions -- fifteen times longer than the characteristic shear-layer oscillation -- are computed to approximate long-time statistical quantities. The SPOD modes capture key dynamical features effectively, such as the interaction between the dominant vortex and corner flows, mediated by the shear layer. We evaluated two optimisation strategies: one that fixes the fundamental frequency and maintains consistency between the model and the modes, and one that relaxes this consistency and treats the fundamental frequency as an additional optimisation variable. Our results indicate that optimising the fundamental frequency is crucial; otherwise, the ROM residuals remain nonzero due to the non-vanishing gradient with respect to this variable. The optimisation process leads to both {zero and non-zero} minima, with only the {zero-}minima producing valid ROM solutions, i.e.~those that drive the objective function and residuals to zero. We also proposed two different protocols to generate initial guesses, viz.~(A) from projection of the modes to DNS data and (B) from random Gaussian sampling based on these projections. {Both protocols derive from the DNS data, albeit to different degrees, and although} multiple unique periodic solutions are identified, only those obtained using protocol A reproduce the essential dynamics of the dominant vortex and shear layers with the correct amplitude. {This suggests that the initialisation plays a key role in guiding the optimisation, an effect likely exacerbated by the strong non-convexity of the problem and the large number of degrees of freedom.}

These solutions also produce good estimates for the statistical distribution of several turbulence quantities. In particular, this is achieved without the use of calibration or closure methods, in contrast to traditional space-only ROMs that often require such corrections to predict flow features with the right amplitude~\citep{ahmed_closures_2021}.

Despite these promising results, there are open questions that suggest several avenues for further investigation. First, the ROM tends to overpredict the energy at frequency–mode pairs near the truncation boundary. This leads to spatio-temporal features in the vorticity field that exhibit higher amplitudes than expected. It may be argued that modal truncation disrupts the natural transfer of energy between spatio-temporal scales, and thus the mechanism responsible for small-scale energy dissipation in the full Navier–Stokes system is not preserved. A more detailed analysis of the energy exchange between resolved and truncated space–time scales is needed, to guide the development of novel closure techniques, {modelling the temporal truncation}, that enhance the accuracy of energy distribution predictions across scales.

Second, the approach leads to models with a large number of amplitude coefficients to be determined, since the entire space–time trajectory is encoded. {The formulation is, however, directly applicable to three-dimensional broadband flows.} Encouragingly, the results obtained {here} with models retaining only 90\% of the turbulent kinetic energy suggest that even highly truncated (and smaller) space–time models may still be relatively robust. {When scaling to larger three-dimensional systems, the principal limitation we foresee is} the cost of evaluating the quadratic term using the convolution sum in \eqref{equ:NS_incomp_SPOD_proj_model} { -- which scales as $O(M^3 N^2)$ with the number of retained modes $M$ and frequencies $N$}. Therefore, alternative strategies, as advocated in \citet{frame_space-time_2024}, should be explored {to reduce this computational burden and enable applications to more complex, practically relevant turbulent flows}. This motivates the investigation of nonlinear couplings and the sparsity of triadic interactions \citep{rubini_l_2020,frame_space-time_2024} between space–time modes. For shear flows, the sparsity of the frequency–wavenumber spectrum \citep{DEL_alamo} indicates that significant computational savings may be achieved by retaining only selected temporal frequencies and spatial wavenumbers where energy is concentrated{, which aligns with the aforementioned objective of efficient three-dimensional flow deployment}.

{Third, we have analysed solutions over a single, albeit long, time interval -- fifteen times longer than the period of the tonal cavity behaviour. Clearly, the expectation is that increasing the time horizon $T$ would improve the frequency resolution and allow access to lower frequencies, where significant energy may still be concentrated. However, an open question is how the method’s predictions and performance vary with shorter or longer time horizons. If the objective is to approximate the statistics of long-time DNS solutions, one should assess the convergence of ROM statistics as $T$ increases. For long UPOs of low-dimensional systems, we have previously shown \citep{lasagna_sensitivity_2020} that convergence occurs at the rate predicted by the central limit theorem, i.e. as $T^{-1/2}$, provided correlations decay sufficiently fast –– typically exponentially. This implies that, asymptotically, longer periodic orbits yield more accurate estimates of long-time properties. Whether the same behaviour holds for ROM solutions remains to be established.}

{Lastly, an important question is whether solutions of the ROM, representing time-periodic flow behaviour developing in the low-order subspace, should be considered to be approximations of, or lie close to, exact time-periodic solutions of the Navier–Stokes equations, i.e.~UPOs, embedded in the high-dimensional chaotic attractor. More broadly, it may be important to establish a more solid link between these two areas of research. We suggest a first step in this direction, left for future work. In principle, considering expansion \eqref{equ:SPOD_finite_truncated_mean}, if the space-time basis was complete (e.g.~using resolvent modes rather than SPOD modes) and both $M$ and $N$ were sufficiently large, one would expect the ROM solutions to correspond to those of the Navier-Stokes equations (or, in the limit, the spatially discretized equations), as in \citet{Burton2025ResolventBased}. It may be possible to set up a continuation-like analysis on these solutions where the number of retained modes is progressively decreased, leading to a bifurcation plot that may shed some light on the link between exact solutions and ROM solutions. The nonlinear nature of the problem suggests that this continuation analysis may involve bifurcations, e.g.~turning points. For instance, as the number of retained modes is reduced, an exact solution may disappear below a critical truncation level, or conversely, a ROM solution may disappear when the basis is refined. This investigation may be feasible in a lower-Reynolds-number flow where UPOs of the full-order system can be computed explicitly. In the present case, focused on a high-Reynolds-number flow where UPOs are not accessible, this experiment is out of reach, yet ROM solutions still approximate the statistical properties of chaotic trajectories. This implies that, at least in the present case, the ROM solution lie in a region of state space within the chaotic attractor visited by turbulence and may thus be used as an efficient and tractable means to approximate the invariant measure, especially when the reduced basis captures the most energetic structures.}

\backsection[Declaration of interests]{The authors report no conflict of interest.}




\appendix

{
\section{Model coefficients in the space-time ROM}\label{app:model_coeffs}
The space-time ROM is built by substituting the equation \eqref{equ:SPOD_finite_truncated_mean} into the NS equations \eqref{equ:NS_incomp} and then projecting the governing equations onto the space-time basis functions $\bm{\phi}_{m}^{l}(\bm{x}) \, e^{i l \omega t}$ using the space-time inner product in the equation \eqref{equ:space_time_inner_product}. This Galerkin projection yields 
\begin{equation}
\begin{aligned}
&\left[ \bm{\phi}_{m}^{l} e^{i l \omega t},
\frac{\partial \left( \bm{U} + \bm{u}'_{r} + \bm{u}'_{u} \right) }{\partial t} + \left( \bm{U} + \bm{u}'_{r} + \bm{u}'_{u} \right) \cdot \nabla \left( \bm{U} + \bm{u}'_{r} + \bm{u}'_{u} \right) \right.
\\ &
\left. - \frac{1}{Re} \nabla \cdot \nabla \left( \bm{U} + \bm{u}'_{r} + \bm{u}'_{u} \right)  \right] = - \left[ \bm{\phi}_{m}^{l} e^{i l \omega t},  \, \nabla p \right]
\, .
\end{aligned}
\label{equ:space_time_Galerkin_projection_expression}
\end{equation}
The projection involved in the unresolved velocity $\bm{u}'_{u}(\bm{x}, t)$ leads to the term $G_m^l$, which is derived in \cref{app:unresolved_interaction}, and the right-hand side related to pressure gradients results in the term $P_m^l$, which is formulated in \cref{app:pressure_gradient}.
Here, we retain $M$ SPOD modes at wavenumbers $k \in [-N, N]$ for the resolved velocity fluctuation field $\bm{u}'_{r}(\bm{x}, t)$ and derive the model coefficient $\bm{L}, \bm{Q}$ and $\bm{C}$ as follows. The left-hand side in \eqref{equ:space_time_Galerkin_projection_expression} is expressed as
\begin{equation}
\begin{aligned}
&\left[ \bm{\phi}_{m}^{l} e^{i l \omega t},  \,  \frac{\partial }{\partial t} \left( \bm{U} + \sum_{n=1}^{M} \sum_{k=-N}^{N} a_{n}^{k} \bm{\phi}_{n}^{k} e^{i k \omega t} \right)  \right]\\
+&
\left[\bm{\phi}_{m}^{l} e^{i l \omega t},  \, \left( \bm{U} +  \sum_{n=1}^{M} \sum_{k=-N}^{N}  a_{n}^{k} \bm{\phi}_{n}^{k} e^{i k \omega t} \right) \cdot \nabla   \left( \bm{U} + \sum_{p=1}^{M} \sum_{q=-N}^{N} a_{p}^{q}  \bm{\phi}_{p}^{q} e^{i q \omega t} \right)  \right] \\
+& 
\left[ \bm{\phi}_{m}^{l} e^{i l \omega t}, \, -\frac{1}{Re} \nabla \cdot \nabla \left( \bm{U} + \sum_{n=1}^{M} \sum_{k=-N}^{N} a_{n}^{k} \bm{\phi}_{n}^{k} e^{i k \omega t} \right) \right] 
+ T G_m^l
\\
=
& \sum_{n=1}^{M} \sum_{k=-N}^{N} i k \omega a_{n}^{k} \left( \bm{\phi}_{m}^{l}, \, \bm{\phi}_{n}^{k} \right) \left( e^{i l \omega t}, \,  e^{i k \omega t} \right)_{T}
+
\left( \bm{\phi}_{m}^{l}, \,   \bm{U} \cdot \nabla \bm{U} \right) \left( e^{i l \omega t}, \, 1 \right)_{T} \\ 
+ &
\sum_{p=1}^{M} \sum_{q=-N}^{N} a_{p}^{q} \left( \bm{\phi}_{m}^{l}, \,  \bm{U} \cdot \nabla \bm{\phi}_{p}^{q} \right) \left( e^{i l \omega t}, \, e^{i q \omega t}  \right)_{T} \\ 
+& \sum_{n=1}^{M} \sum_{k=-N}^{N} a_{n}^{k} \left( \bm{\phi}_{m}^{l}, \, \bm{\phi}_{n}^{k} \cdot  \nabla \bm{U} \right) \left( e^{i l \omega t}, \,  e^{i k \omega t} \right)_{T} \\
+&
\sum_{n=1}^{M} \sum_{p=1}^{M} \sum_{k=-N}^{N} \sum_{q=-N}^{N} a_{n}^{k}  a_{p}^{q} \left( \bm{\phi}_{m}^{l}, \, \bm{\phi}_{n}^{k} \cdot \nabla  \bm{\phi}_{p}^{q} \right) \left( e^{i l \omega t}, \, e^{i (k+q) \omega t}\right)_{T}
\\
+& \sum_{n=1}^{M} \sum_{k=-N}^{N} a_{n}^{k} \left(-\frac{1}{Re}\right) \left( \bm{\phi}_{m}^{l}, \,  \nabla \cdot \nabla \bm{\phi}_{n}^{k} \right) \left( e^{i l \omega t}, \,   e^{i k \omega t} \right)_{T}
\\
- &\frac{1}{Re} \left( \bm{\phi}_{m}^{l}, \,  \nabla \cdot \nabla \bm{U} \right) \left( e^{i l \omega t}, \,  1 \right)_{T} 
+ T G_m^l
\, ,
\end{aligned}
\label{equ:NS_incomp_SPOD_rmMean_time_proj_M}
\end{equation}
where the temporal inner product $\left( f, \, g \right)_{T}$ is defined as
\begin{equation}
 \left( f, \, g \right)_{T} = \int_{t=0}^{T} \bar{f} \, g \, \mathrm{d}t
 \, .
\end{equation}
The orthogonality of Fourier modes implies that
\begin{equation}
\begin{aligned}
\left( e^{i l \omega t}, \,  e^{i k \omega t} \right)_{T} &= \delta_{l,k} T
\, ,
\end{aligned}
\end{equation}
where $\delta_{l,k}$ denotes the Kronecker delta.
Therefore, the expression in \eqref{equ:NS_incomp_SPOD_rmMean_time_proj_M} is equivalent to
\begin{equation}
\begin{aligned}
%
& T \sum_{n=1}^{M} a_{n}^{l} i l \omega \left( \bm{\phi}_{m}^{l}, \, \bm{\phi}_{n}^{l} \right)
+ T \delta_{l,0} \left( \bm{\phi}_{m}^{l}, \, \bm{U} \cdot \nabla \bm{U} \right) 
+ T \sum_{p=1}^{M} a_{p}^{l} \left( \bm{\phi}_{m}^{l}, \,  \bm{U} \cdot \nabla \bm{\phi}_{p}^{l} \right) \\
+& T \sum_{n=1}^{M} a_{n}^{l} \left( \bm{\phi}_{m}^{l}, \, \bm{\phi}_{n}^{l} \cdot  \nabla \bm{U} \right)  
+ T \sum_{n=1}^{M} \sum_{p=1}^{M} \sum_{k=-N}^{N} a_{n}^{k}  a_{p}^{l-k} \left( \bm{\phi}_{m}^{l}, \, \bm{\phi}_{n}^{k} \cdot \nabla  \bm{\phi}_{p}^{l-k} \right) \\
- & T \delta_{l,0} \frac{1}{Re} \left( \bm{\phi}_{m}^{l}, \,  \nabla \cdot \nabla \bm{U} \right)
+ T \sum_{n=1}^{M} a_{n}^{l} \left(-\frac{1}{Re}\right) \left( \bm{\phi}_{m}^{l}, \,  \nabla \cdot \nabla \bm{\phi}_{n}^{l} \right)
+ T G_m^l
\, .
\end{aligned}
\label{equ:NS_incomp_projection_1}
\end{equation}
Given that the right-hand side leads to $- T P_m^l$, rearranging and simplifying the terms in \eqref{equ:NS_incomp_projection_1} results in 
\begin{equation}
\begin{aligned}
& \sum_{n=1}^{M} a_{n}^{l} 
\left( 
i l \omega \delta_{m,n} 
+ \left( \bm{\phi}_{m}^{l}, \, \bm{U} \cdot \nabla \bm{\phi}_{n}^{l} \right) 
+ \left( \bm{\phi}_{m}^{l}, \, \bm{\phi}_{n}^{l} \cdot \nabla \bm{U} \right) 
-\frac{1}{Re} \left( \bm{\phi}_{m}^{l}, \,  \nabla \cdot \nabla \bm{\phi}_{n}^{l} \right)
\right) \\
+& \sum_{n=1}^{M} \sum_{p=1}^{M} \sum_{k=-N}^{N} a_{n}^{k}  a_{p}^{l-k} \left( \bm{\phi}_{m}^{l}, \, \bm{\phi}_{n}^{k} \cdot \nabla  \bm{\phi}_{p}^{l-k} \right) 
- \delta_{l,0} \frac{1}{Re} \left( \bm{\phi}_{m}^{l}, \,  \nabla \cdot \nabla \bm{U} \right)
+ \delta_{l,0} \left( \bm{\phi}_{m}^{l}, \, \bm{U} \cdot \nabla \bm{U} \right)
\\
+& G_m^l = - P_m^l
\, ,
\end{aligned}
\label{equ:NS_incomp_projection_2}
\end{equation}
where $\left( \bm{\phi}_{m}^{l}, \, \bm{\phi}_{n}^{l} \right) = \delta_{m,n}$ due to the orthogonality of SPOD modes at the same frequency.
This yields a system of nonlinear algebraic equations as shown in \eqref{equ:NS_incomp_SPOD_proj_model},
where the model coefficients of $\bm{L}, \bm{Q}$ and $\bm{C}$ are obtained as defined in the equations \eqref{equ:model_coeffs_expression}.
}

\section{Unresolved interactions}\label{app:unresolved_interaction}
The velocity decomposition of the euquation \eqref{equ:SPOD_finite_truncated_mean} contains resolved $\bm{u}'_{r}(\bm{x}, t)$ and unresolved components $\bm{u}'_{u}(\bm{x}, t)$. Substituting this decomposition into the Navier--Stokes equations and using Galerkin projection onto each of the space-time basis functions $\bm{\phi}_m^l(\bm{x}) e^{i l \omega t} $ yields several terms containing the unresolved components lumped together as
\begin{equation}
\begin{aligned}
G_m^l &= \frac{1}{T} \left[\bm{\phi}_m^l e^{i l \omega t} , \,  \frac{\partial \bm{u}'_{u} }{\partial t} \right]
+ \frac{1}{T} \left[\bm{\phi}_m^l e^{i l \omega t} , \,  { \left(\bm{U} \cdot \nabla \right) \bm{u}'_{u} +  \left(\bm{u}'_{u} \cdot \nabla \right) \bm{U}} \right] \\
&- \frac{1}{T} \left[ \bm{\phi}_m^l e^{i l \omega t} , \,  \frac{1}{Re} \nabla \cdot \nabla \bm{u}'_{u} \right] \\
&+ \frac{1}{T} \left[ \bm{\phi}_m^l e^{i l \omega t} ,  \,  {\left( \bm{u}'_{r}  \cdot \nabla \right) \bm{u}'_{u} + \left( \bm{u}'_{u}  \cdot \nabla \right)  \bm{u}'_{r}  +  \left( \bm{u}'_{u}  \cdot \nabla \right) \bm{u}'_{u} } \right]
\, .
\end{aligned}
\label{equ:T}
\end{equation}
There are linear and nonlinear contributions to the unresolved interactions $G_m^l$. It is noted that, due to the orthogonality of Fourier modes, only the frequencies corresponding to the fundamental frequency contribute to these unresolved interactions.

{
\section{Pressure gradient term in the space-time ROM}\label{app:pressure_gradient}
The projection of the pressure gradient term onto the space-time basis function $\bm{\phi}_{m}^{l}(\bm{x}) e^{i l \omega t}$ read as
\begin{equation}
\begin{aligned}
P_m^l &= \frac{1}{T} \left[\bm{\phi}_{m}^{l} e^{i l \omega t} ,  \, \nabla p \right] \\
&= \frac{1}{T} \int_{0}^{T} \overline{e^{i l \omega t}}  \int_{V}  \overline{\bm{\phi}_{m}^{l}} \cdot \nabla p \, \mathrm{d}V   \, \mathrm{d}t\\
&= \frac{1}{T} \int_{0}^{T} \overline{e^{i l \omega t}} \left( \int_{\Gamma} p \, \overline{\bm{\phi}_{m}^{l}} \cdot \bm{n} \, \mathrm{d}\Gamma - \int_{V}  p \nabla \cdot \overline{\bm{\phi}_{m}^{l}} \, \mathrm{d}V \right)  \, \mathrm{d}t
\, ,
\end{aligned}
\label{equ:NS_incomp_SPOD_pressure}
\end{equation}
where $\Gamma$ is the boundary of the domain $V$. The volume integral in the brackets vanishes when the spatial basis functions are divergence free, viz.~$\nabla \cdot \bm{\phi}_{m}^{l} = 0$, while the boundary term vanishes for some boundary conditions, e.g.~periodic or no-slip conditions.}

\bibliographystyle{jfm}
\bibliography{jfm}

@article{Schlegel_Noack_2015, title={On long-term boundedness of {G}alerkin models}, volume={765}, DOI={10.1017/jfm.2014.736}, journal={J. Fluid Mech.}, author={Schlegel, M. and Noack, B. R.}, year={2015}, pages={325–352}}

@article{beck,
  title = {Data-driven guessing and gluing of unstable periodic orbits},
  author = {Beck, P. and Parker, J. P. and Schneider, T. M.},
  journal = {Phys. Rev. E},
  volume = {112},
  issue = {2},
  pages = {024203},
  numpages = {18},
  year = {2025},
  month = {Aug},
  publisher = {American Physical Society},
  doi = {10.1103/9vjc-g86s},
  url = {https://link.aps.org/doi/10.1103/9vjc-g86s}
}

@article{Budanur2015,
  author    = {Budanur, N. B. and Cvitanovi{\'c}, P. and Davidchack, R. L. and Siminos, E.},
  title     = {Reduction of {SO}(2) Symmetry for Spatially Extended Dynamical Systems},
  journal   = {Phys. Rev. Lett.},
  volume    = {114},
  pages     = {084102},
  year      = {2015},
  doi       = {10.1103/PhysRevLett.114.084102}
}

@article{Citro,
	title = {Adjoint-based sensitivity analysis of periodic orbits by the Fourier–Galerkin method},
	journal = {J. Comput. Phys.},
	volume = {440},
	pages = {110403},
	year = {2021},
	issn = {0021-9991},
	doi = {https://doi.org/10.1016/j.jcp.2021.110403},
	url = {https://www.sciencedirect.com/science/article/pii/S0021999121002989},
	author = {J. Sierra and P. Jolivet and F. Giannetti and V. Citro},
}

@article{Bergmann,
    author = {Bergmann, M. and Cordier, L. and Brancher, J.P.},
    title = {Optimal rotary control of the cylinder wake using proper orthogonal decomposition reduced-order model},
    journal = {Phys. Fluids},
    volume = {17},
    number = {9},
    pages = {097101},
    year = {2005},
    month = {08},
    issn = {1070-6631},
    doi = {10.1063/1.2033624}
}

@article{Maia_Cavalieri_2025,
	title={Turbulence suppression in plane {C}ouette flow using reduced-order models},
	volume={1014},
	DOI={10.1017/jfm.2025.10258},
	journal={J. Fluid Mech.},
	author={Maia, I. A. and Cavalieri, A.}, 
	year={2025}, 
	pages={A18}
}

@article{PADOVAN2024112597,
	title = {Continuous-time balanced truncation for time-periodic fluid flows using frequential Gramians},
	journal = {J. Comput. Phys.},
	volume = {496},
	pages = {112597},
	year = {2024},
	issn = {0021-9991},
	author = {Padovan, P. and Rowley, C. W.}
}

@article{Burton2025ResolventBased,
  author    = {Burton, T. and Symon, S. and Lasagna, D.},
  title     = {Resolvent-Based Optimisation for Invariant Flows},
  journal   = {arXiv preprint},
  volume    = {arXiv:2509.02463},
  year      = {2025},
  eprint    = {2509.02463},
  archivePrefix = {arXiv},
  primaryClass  = {physics.flu-dyn}
}

@article{SIERRAAUSIN2022114736,
	title = {Efficient computation of time-periodic compressible flows with spectral techniques},
	journal = {Comput. Methods Appl. Mech. Engrg.},
	volume = {393},
	pages = {114736},
	year = {2022},
	issn = {0045-7825},
	doi = {https://doi.org/10.1016/j.cma.2022.114736},
	url = {https://www.sciencedirect.com/science/article/pii/S0045782522000949},
	author = {Sierra-Ausin, J. and Citro, V. and Giannetti, F. and Fabre, D.}
}

@article{Rigas_Sipp_Colonius_2021,
	title={Nonlinear input/output analysis: application to boundary layer transition},
	volume={911}, 
	DOI={10.1017/jfm.2020.982}, 
	journal={J. Fluid Mech.}, 
	author={Rigas, G. and Sipp, D. and Colonius, T.},
	year={2021},
	pages={A15}
}

@article{Chandler_Kerswell_2013, 
	title={Invariant recurrent solutions embedded in a turbulent two-dimensional Kolmogorov flow}, 
	volume={722}, 
	DOI={10.1017/jfm.2013.122}, 
	journal={J. Fluid Mech.}, 
	author={Chandler, G. J. and Kerswell, R. R.}, 
	year={2013}, 
	pages={554–595}
}

@article{PICARD2000359,
	title = {Pressure velocity coupling in a subsonic round jet},
	journal = {Int. J. Heat Fluid Fl.},
	volume = {21},
	number = {3},
	pages = {359-364},
	year = {2000},
	issn = {0142-727X},
	author = {Picard, C. and Delville, J.},
}

@article{peng,
    author = {Peng, M. and Kaptanoglu, A. A. and Hansen, C. J. and Stevens-Haas, J. and Manohar, K. and Brunton, S. L.},
    title = {Extending the trapping theorem to provide local stability guarantees for quadratically nonlinear models},
    journal = {Phys. Fluids},
    volume = {37},
    number = {10},
    pages = {107115},
    year = {2025},
    month = {10},
    issn = {1070-6631},
    doi = {10.1063/5.0287432}
}

@article{zheng2024ghost,
	title = {Ghost states underlying spatial and temporal patterns: How nonexistent invariant solutions control nonlinear dynamics},
	author = {Zheng, Z. and Beck, P. and Yang, T. and Ashtari, O. and Parker, J. P. and Schneider, T. M.},
	journal = {Phys. Rev. E},
	volume = {112},
	issue = {2},
	pages = {024212},
	numpages = {19},
	year = {2025},
	month = {Aug},
	publisher = {American Physical Society},
	doi = {10.1103/8b86-3plx},
	url = {https://link.aps.org/doi/10.1103/8b86-3plx}
}

@article{artuso1990recycling,
	title={Recycling of strange sets: I. Cycle expansions},
	author={Artuso, R. and Aurell, E. and Cvitanovic, P.},
	journal={Nonlinearity},
	volume={3},
	number={2},
	pages={325},
	year={1990},
	publisher={IOP Publishing}
}

@article{Wang_Ayats_Deguchi_Meseguer_Mellibovsky_2025, 
	title={Mathematically established chaos and forecast of statistics with recurrent patterns in Taylor–Couette flow}, 
	volume={1011}, 
	DOI={10.1017/jfm.2025.151}, 
	journal={J. Fluid Mech.}, 
	author={Wang, B. and Ayats, R. and Deguchi, K. and Meseguer, A. and Mellibovsky, F.}, 
	year={2025}, 
	pages={R2}
}

@article{DEL_alamo, 
    title={Estimation of turbulent convection velocities and corrections to {T}aylor’s approximation}, 
    volume={640},
    DOI={10.1017/S0022112009991029}, 
    journal={J. Fluid Mech.}, 
    author={Del \'Alamo, J. C. and Jim\'enez, J.}, 
    year={2009}, 
    pages={5–26}
}

@article{Cavalieri,
  title = {Reduced-order {G}alerkin models of plane {C}ouette flow},
  author = {Cavalieri, A. V. G. and Nogueira, P. A. S.},
  journal = {Phys. Rev. Fluids},
  volume = {7},
  issue = {10},
  pages = {L102601},
  numpages = {10},
  year = {2022},
  month = {Oct},
  publisher = {American Physical Society},
  doi = {10.1103/PhysRevFluids.7.L102601},
  url = {https://link.aps.org/doi/10.1103/PhysRevFluids.7.L102601}
}

@article{hopf_mathematical_1948,
	title = {A mathematical example displaying features of turbulence},
	volume = {1},
	issn = {1097-0312},
	doi = {10.1002/cpa.3160010401},
	language = {en},
	number = {4},
	urldate = {2024-04-15},
	journal = {Commun. Pure Appl. Math.},
	author = {Hopf, E.},
	year = {1948},
	pages = {303--322}
}

@book{lumley_stochastic_1970,
	address = {New York},
	series = {Applied mathematics and mechanics},
	title = {Stochastic {Tools} in {Turbulence}},
	language = {English},
	urldate = {2021-04-01},
	publisher = {Academic Press},
	author = {Lumley, J. L.},
    number = {},
	year = {1970}
}

@article{cazemier_proper_1998,
	title = {Proper orthogonal decomposition and low-dimensional models for driven cavity flows},
	volume = {10},
	issn = {1070-6631},
	url = {https://doi.org/10.1063/1.869686},
	doi = {10.1063/1.869686},
	number = {7},
	journal = {Phys. Fluids},
	author = {Cazemier, W. and Verstappen, R. W. C. P. and Veldman, A. E. P.},
	month = jul,
	year = {1998},
	pages = {1685--1699}
}

@article{he_efficient_1998,
	title = {Efficient {Approach} for {Analysis} of {Unsteady} {Viscous} {Flows} in {Turbomachines}},
	volume = {36},
	issn = {0001-1452, 1533-385X},
	url = {https://arc.aiaa.org/doi/10.2514/2.328},
	doi = {10.2514/2.328},
	language = {en},
	number = {11},
	urldate = {2022-10-07},
	journal = {AIAA J.},
	author = {He, L. and Ning, W.},
	month = nov,
	year = {1998},
	pages = {2005--2012}
}

@article{davidchack_efficient_1999,
	title = {Efficient algorithm for detecting unstable periodic orbits in chaotic systems},
	volume = {60},
	copyright = {http://link.aps.org/licenses/aps-default-license},
	issn = {1063-651X, 1095-3787},
	url = {https://link.aps.org/doi/10.1103/PhysRevE.60.6172},
	doi = {10.1103/PhysRevE.60.6172},
	language = {en},
	number = {5},
	urldate = {2025-06-26},
	journal = {Phys. Rev. E},
	author = {Davidchack, R. L. and Lai, Y.},
	month = nov,
	year = {1999},
	pages = {6172--6175}
}

@article{lea_sensitivity_2000,
	title = {Sensitivity analysis of the climate of a chaotic system},
	volume = {52},
	issn = {0280-6495, 1600-0870},
	url = {http://tellusa.net/index.php/tellusa/article/view/12283},
	doi = {10.1034/j.1600-0870.2000.01137.x},
	language = {en},
	number = {5},
	urldate = {2018-11-15},
	journal = {Tellus A},
	author = {Lea, Daniel J. and Allen, Myles R. and Haine, Thomas W. N.},
	month = oct,
	year = {2000},
	pages = {523--532}
}

@article{kawahara_periodic_2001,
	title = {Periodic motion embedded in plane {Couette} turbulence: regeneration cycle and burst},
	volume = {449},
	issn = {0022-1120, 1469-7645},
	shorttitle = {Periodic motion embedded in plane {Couette} turbulence},
	url = {https://www.cambridge.org/core/product/identifier/S0022112001006243/type/journal_article},
	doi = {10.1017/S0022112001006243},
	urldate = {2022-11-29},
	journal = {J. Fluid Mech.},
	author = {Kawahara, G. and Kida, S.},
	month = dec,
	year = {2001},
	pages = {291--300}
}

@article{hall_computation_2002,
	title = {Computation of {Unsteady} {Nonlinear} {Flows} in {Cascades} {Using} a {Harmonic} {Balance} {Technique}},
	volume = {40},
	issn = {0001-1452, 1533-385X},
	url = {https://arc.aiaa.org/doi/10.2514/2.1754},
	doi = {10.2514/2.1754},
	language = {en},
	number = {5},
	urldate = {2022-09-30},
	journal = {AIAA J.},
	author = {Hall, K. C. and Thomas, J. P. and Clark, W. S.},
	month = may,
	year = {2002},
	pages = {879--886}
}

@article{noack_hierarchy_2003,
	title = {A hierarchy of low-dimensional models for the transient and post-transient cylinder wake},
	volume = {497},
	issn = {00221120, 14697645},
	url = {http://www.journals.cambridge.org/abstract_S0022112003006694},
	doi = {10.1017/S0022112003006694},
	language = {en},
	urldate = {2019-08-05},
	journal = {J. Fluid Mech.},
	author = {Noack, B. R. and Afanasiev, K. and Morzy\'{n}ski, M. and Tadmor, G. and Thiele, F.},
	month = dec,
	year = {2003},
	pages = {335--363}
}

@article{hof_experimental_2004,
	title = {Experimental {Observation} of {Nonlinear} {Traveling} {Waves} in {Turbulent} {Pipe} {Flow}},
	volume = {305},
	issn = {0036-8075},
	url = {https://www.jstor.org/stable/3837916},
	number = {5690},
	urldate = {2024-09-18},
	journal = {Science},
	author = {Hof, B. and van Doorne, C. W. H. and Westerweel, J. and Nieuwstadt, F. T. M. and Faisst, H. and Eckhardt, B. and Wedin, H. and Kerswell, R. R. and Waleffe, F.},
	year = {2004},
	pages = {1594--1598}
}

@article{rowley_model_2004,
	title = {Model reduction for compressible flows using {POD} and {Galerkin} projection},
	volume = {189},
	issn = {0167-2789},
	url = {https://www.sciencedirect.com/science/article/pii/S0167278903003841},
	doi = {10.1016/j.physd.2003.03.001},
	number = {1},
	urldate = {2025-05-14},
	journal = {Physica D.},
	author = {Rowley, C. W. and Colonius, T. and Murray, R. M.},
	month = feb,
	year = {2004},
	pages = {115--129}
}

@article{couplet_calibrated_2005,
	title = {Calibrated reduced-order {POD}-{Galerkin} system for fluid flow modelling},
	volume = {207},
	copyright = {https://www.elsevier.com/tdm/userlicense/1.0/},
	issn = {00219991},
	url = {https://linkinghub.elsevier.com/retrieve/pii/S0021999105000239},
	doi = {10.1016/j.jcp.2005.01.008},
	language = {en},
	number = {1},
	urldate = {2025-04-14},
	journal = {J. Comput. Phys.},
	author = {Couplet, M. and Basdevant, C. and Sagaut, P.},
	month = jul,
	year = {2005},
	pages = {192--220}
}

@article{jovanovic_componentwise_2005,
	title = {Componentwise energy amplification in channel flows},
	volume = {534},
	issn = {0022-1120, 1469-7645},
	url = {http://www.journals.cambridge.org/abstract_S0022112005004295},
	doi = {10.1017/S0022112005004295},
	language = {en},
	urldate = {2024-09-23},
	journal = {J. Fluid Mech.},
	author = {Jovanovi\'{c}, M. R. and Bamieh, B.},
	month = jun,
	year = {2005},
	pages = {145--183}
}

@article{barbagallo_closed-loop_2009,
	title = {Closed-loop control of an open cavity flow using reduced-order models},
	volume = {641},
	issn = {1469-7645, 0022-1120},
	url = {https://www.cambridge.org/core/journals/journal-of-fluid-mechanics/article/closedloop-control-of-an-open-cavity-flow-using-reducedorder-models/D266AB7BA9E4868A2C6F8DB809CAB32E},
	doi = {10.1017/S0022112009991418},
	language = {en},
	urldate = {2025-04-13},
	journal = {J. Fluid Mech.},
	author = {Barbagallo, A. and Sipp, D. and Schmid, P. J.},
	month = dec,
	year = {2009},
	pages = {1--50}
}

@article{saiki_time-averaged_2009,
	title = {Time-averaged properties of unstable periodic orbits and chaotic orbits in ordinary differential equation systems},
	volume = {79},
	copyright = {http://link.aps.org/licenses/aps-default-license},
	issn = {1539-3755, 1550-2376},
	url = {https://link.aps.org/doi/10.1103/PhysRevE.79.015201},
	doi = {10.1103/PhysRevE.79.015201},
	language = {en},
	number = {1},
	urldate = {2024-09-20},
	journal = {Phys. Rev. E},
	author = {Saiki, Y. and Yamada, M.},
	month = jan,
	year = {2009},
	pages = {015201}
}

@article{mckeon_critical-layer_2010,
	title = {A critical-layer framework for turbulent pipe flow},
	volume = {658},
	copyright = {https://www.cambridge.org/core/terms},
	issn = {0022-1120, 1469-7645},
	url = {https://www.cambridge.org/core/product/identifier/S002211201000176X/type/journal_article},
	doi = {10.1017/S002211201000176X},
	language = {en},
	urldate = {2024-09-22},
	journal = {J. Fluid Mech.},
	author = {McKeon, B. J. and Sharma, A. S.},
	month = sep,
	year = {2010},
	pages = {336--382}
}

@article{sipp_dynamics_2010,
	title = {Dynamics and {Control} of {Global} {Instabilities} in {Open}-{Flows}: {A} {Linearized} {Approach}},
	volume = {63},
	issn = {0003-6900, 2379-0407},
	shorttitle = {Dynamics and {Control} of {Global} {Instabilities} in {Open}-{Flows}},
	url = {https://asmedigitalcollection.asme.org/appliedmechanicsreviews/article/doi/10.1115/1.4001478/446466/Dynamics-and-Control-of-Global-Instabilities-in},
	doi = {10.1115/1.4001478},
	language = {en},
	number = {3},
	urldate = {2024-09-23},
	journal = {Appl. Mech. Rev.},
	author = {Sipp, D. and Marquet, O. and Meliga, P. and Barbagallo, A.},
	month = may,
	year = {2010},
	pages = {030801}
}

@article{carlberg_efficient_2011,
	title = {Efficient non‐linear model reduction via a least‐squares {Petrov}–{Galerkin} projection and compressive tensor approximations},
	volume = {86},
	copyright = {http://onlinelibrary.wiley.com/termsAndConditions\#vor},
	issn = {0029-5981, 1097-0207},
	url = {https://onlinelibrary.wiley.com/doi/10.1002/nme.3050},
	doi = {10.1002/nme.3050},
	language = {en},
	number = {2},
	urldate = {2025-05-14},
	journal = {Int. J. Numer. Methods Eng.},
	author = {Carlberg, K. and Bou‐Mosleh, C. and Farhat, C.},
	month = apr,
	year = {2011},
	pages = {155--181}
}

@article{semeraro_feedback_2011,
	title = {Feedback control of three-dimensional optimal disturbances using reduced-order models},
	volume = {677},
	issn = {1469-7645, 0022-1120},
	url = {https://www.cambridge.org/core/journals/journal-of-fluid-mechanics/article/feedback-control-of-threedimensional-optimal-disturbances-using-reducedorder-models/6E4D3242D3B88F0061DFE6C70FE109B4},
	doi = {10.1017/S0022112011000620},
	language = {en},
	urldate = {2025-04-13},
	journal = {J. Fluid Mech.},
	author = {Semeraro, O. and Bagheri, S. and Brandt, L. and Henningson, D. S.},
	month = jun,
	year = {2011},
	pages = {63--102}
}

@article{tadmor_reduced-order_2011,
	title = {Reduced-order models for closed-loop wake control},
	volume = {369},
	url = {https://royalsocietypublishing.org/doi/full/10.1098/rsta.2010.0367},
	doi = {10.1098/rsta.2010.0367},
	number = {1940},
	urldate = {2025-04-13},
	journal = {Philos. Trans. R. Soc. A},
	author = {Tadmor, G. and Lehmann, O. and Noack, B. R. and Cordier, L. and Delville, J. and Bonnet, J.-P. and Morzy\'{n}ski, M.},
	month = apr,
	year = {2011},
	pages = {1513--1524}
}

@book{holmes_turbulence_2012,
	address = {New York},
	edition = {2nd ed},
	series = {Cambridge monographs on mechanics},
	title = {Turbulence, coherent structures, dynamical systems and symmetry},
	isbn = {978-1-107-00825-0},
	url = {https://doi.org/10.1017/CBO9780511622700},
	language = {en},
	number = {},
	publisher = {Cambridge University Press},
	author = {Holmes, P. and Lumey, J. L and Berkooz, G. and Rowley, C. W.},
	year = {2012}
}

@article{kawahara_significance_2012,
	title = {The {Significance} of {Simple} {Invariant} {Solutions} in {Turbulent} {Flows}},
	volume = {44},
	issn = {0066-4189, 1545-4479},
	url = {https://www.annualreviews.org/doi/10.1146/annurev-fluid-120710-101228},
	doi = {10.1146/annurev-fluid-120710-101228},
	language = {en},
	number = {1},
	urldate = {2022-11-29},
	journal = {Annu. Rev. Fluid Mech.},
	author = {Kawahara, G. and Uhlmann, M. and van Veen, L.},
	month = jan,
	year = {2012},
	pages = {203--225}
}

@article{balajewicz_low-dimensional_2013,
	title = {Low-dimensional modelling of high-{Reynolds}-number shear flows incorporating constraints from the {Navier}–{Stokes} equation},
	volume = {729},
	issn = {0022-1120, 1469-7645},
	url = {https://www.cambridge.org/core/product/identifier/S0022112013002784/type/journal_article},
	doi = {10.1017/jfm.2013.278},
	language = {en},
	urldate = {2022-10-25},
	journal = {J. Fluid Mech.},
	author = {Balajewicz, M. J. and Dowell, E. H. and Noack, B. R.},
	month = aug,
	year = {2013},
	pages = {285--308}
}

@article{hall_harmonic_2013,
	title = {Harmonic balance methods applied to computational fluid dynamics problems},
	volume = {27},
	issn = {1061-8562},
	url = {https://doi.org/10.1080/10618562.2012.742512},
	doi = {10.1080/10618562.2012.742512},
	number = {2},
	urldate = {2025-04-14},
	journal = {Int. J. Comput. Fluid Dyn.},
	author = {Hall, K. C. and Ekici, K. and Thomas, J. P. and Dowell, E. H.},
	month = feb,
	year = {2013},
	pages = {52--67}
}

@article{wang_forward_2013,
	title = {Forward and adjoint sensitivity computation of chaotic dynamical systems},
	volume = {235},
	issn = {00219991},
	url = {http://linkinghub.elsevier.com/retrieve/pii/S0021999112005360},
	doi = {10.1016/j.jcp.2012.09.007},
	language = {en},
	urldate = {2018-11-15},
	journal = {J. Comput. Phys.},
	author = {Wang, Q.},
	month = feb,
	year = {2013},
	pages = {1--13}
}

@article{chernyshenko_polynomial_2014,
	title = {Polynomial sum of squares in fluid dynamics: a review with a look ahead},
	volume = {372},
	shorttitle = {Polynomial sum of squares in fluid dynamics},
	url = {https://royalsocietypublishing.org/doi/10.1098/rsta.2013.0350},
	doi = {10.1098/rsta.2013.0350},
	number = {2020},
	urldate = {2025-04-13},
	journal = {Philos. Trans. R. Soc. A},
	author = {Chernyshenko, S. I. and Goulart, P. and Huang, D. and Papachristodoulou, A.},
	month = jul,
	year = {2014},
	pages = {20130350}
}

@article{yano_space-time_2014,
	title = {A {Space}-{Time} {Petrov}--{Galerkin} {Certified} {Reduced} {Basis} {Method}: {Application} to the {Boussinesq} {Equations}},
	volume = {36},
	issn = {1064-8275, 1095-7197},
	shorttitle = {A {Space}-{Time} {Petrov}--{Galerkin} {Certified} {Reduced} {Basis} {Method}},
	url = {http://epubs.siam.org/doi/10.1137/120903300},
	doi = {10.1137/120903300},
	language = {en},
	number = {1},
	urldate = {2025-04-14},
	journal = {SIAM J. Sci. Comput.},
	author = {Yano, M.},
	month = jan,
	year = {2014},
	pages = {A232--A266}
}

@article{farazmand_adjoint-based_2016,
	title = {An adjoint-based approach for finding invariant solutions of {Navier}–{Stokes} equations},
	volume = {795},
	copyright = {https://www.cambridge.org/core/terms},
	issn = {0022-1120, 1469-7645},
	url = {https://www.cambridge.org/core/product/identifier/S0022112016002032/type/journal_article},
	doi = {10.1017/jfm.2016.203},
	language = {en},
	urldate = {2024-09-24},
	journal = {J. Fluid Mech.},
	author = {Farazmand, M.},
	month = may,
	year = {2016},
	pages = {278--312}
}

@article{lasagna_sum--squares_2016,
	title = {Sum-of-squares approach to feedback control of laminar wake flows},
	volume = {809},
	copyright = {http://creativecommons.org/licenses/by/4.0/},
	issn = {0022-1120, 1469-7645},
	url = {https://www.cambridge.org/core/product/identifier/S0022112016006881/type/journal_article},
	doi = {10.1017/jfm.2016.688},
	language = {en},
	urldate = {2025-04-13},
	journal = {J. Fluid Mech.},
	author = {Lasagna, D. and Huang, D. and Tutty, O. R. and Chernyshenko, S.},
	month = dec,
	year = {2016},
	pages = {628--663}
}

@article{sharma_correspondence_2016,
	title = {Correspondence between {Koopman} mode decomposition, resolvent mode decomposition, and invariant solutions of the {Navier}-{Stokes} equations},
	volume = {1},
	url = {https://link.aps.org/doi/10.1103/PhysRevFluids.1.032402},
	doi = {10.1103/PhysRevFluids.1.032402},
	number = {3},
	urldate = {2025-05-02},
	journal = {Phys. Rev. Fluids},
	author = {Sharma, A. S. and Mezi\'{c}, I. and McKeon, B. J.},
	month = jul,
	year = {2016},
	pages = {032402}
}

@article{rowley_model_2017,
	title = {Model reduction for flow analysis and control},
	volume = {49},
	issn = {0066-4189, 1545-4479},
	url = {http://www.annualreviews.org/doi/10.1146/annurev-fluid-010816-060042},
	doi = {10.1146/annurev-fluid-010816-060042},
	language = {en},
	number = {1},
	urldate = {2021-03-30},
	journal = {Annu. Rev. Fluid Mech.},
	author = {Rowley, C. W. and Dawson, S. T. M.},
	month = jan,
	year = {2017},
	pages = {387--417}
}

@article{taira_modal_2017,
	title = {Modal analysis of fluid flows: an overview},
	volume = {55},
	issn = {0001-1452, 1533-385X},
	shorttitle = {Modal {Analysis} of {Fluid} {Flows}},
	url = {https://arc.aiaa.org/doi/10.2514/1.J056060},
	doi = {10.2514/1.J056060},
	language = {en},
	number = {12},
	urldate = {2019-07-26},
	journal = {AIAA J.},
	author = {Taira, K. and Brunton, S. L. and Dawson, S. T. M. and Rowley, C. W. and Colonius, T. and McKeon, B. J. and Schmidt, O. T. and Gordeyev, S. and Theofilis, V. and Ukeiley, L. S.},
	month = dec,
	year = {2017},
	pages = {4013--4041}
}

@article{baumann_space-time_2018,
	title = {Space-{Time} {Galerkin} {POD} with {Application} in {Optimal} {Control} of {Semilinear} {Partial} {Differential} {Equations}},
	volume = {40},
	issn = {1064-8275, 1095-7197},
	url = {https://epubs.siam.org/doi/10.1137/17M1135281},
	doi = {10.1137/17M1135281},
	language = {en},
	number = {3},
	journal = {SIAM J. Sci. Comput.},
	author = {Baumann, M. and Benner, P. and Heiland, J.},
	month = jan,
	year = {2018},
	pages = {A1611--A1641}
}

@article{lasagna_sensitivity_2018,
	title = {Sensitivity {Analysis} of {Chaotic} {Systems} {Using} {Unstable} {Periodic} {Orbits}},
	volume = {17},
	issn = {1536-0040},
	url = {https://epubs.siam.org/doi/10.1137/17M114354X},
	doi = {10.1137/17M114354X},
	language = {en},
	number = {1},
	urldate = {2021-10-20},
	journal = {SIAM J. Appl. Dyn. Syst.},
	author = {Lasagna, D.},
	month = jan,
	year = {2018},
	pages = {547--580}
}

@article{loiseau_constrained_2018,
	title = {Constrained sparse {Galerkin} regression},
	volume = {838},
	issn = {0022-1120, 1469-7645},
	url = {https://www.cambridge.org/core/journals/journal-of-fluid-mechanics/article/constrained-sparse-galerkin-regression/0E18A4A55FF5AC1401D236C0E4D1CAAE},
	doi = {10.1017/jfm.2017.823},
	language = {en},
	urldate = {2023-09-28},
	journal = {J. Fluid Mech.},
	author = {Loiseau, J.-C. and Brunton, S. L.},
	month = mar,
	year = {2018},
	pages = {42--67}
}

@article{mogensen_optim_2018,
	title = {Optim: {A} mathematical optimization package for {Julia}},
	volume = {3},
	issn = {2475-9066},
	shorttitle = {Optim},
	url = {https://joss.theoj.org/papers/10.21105/joss.00615},
	doi = {10.21105/joss.00615},
	language = {en},
	number = {24},
	urldate = {2024-09-12},
	journal = {J. Open Source Softw.},
	author = {Mogensen, P. K. and Riseth, A. N.},
	month = apr,
	year = {2018},
	pages = {615}
}

@article{towne_spectral_2018,
	title = {Spectral proper orthogonal decomposition and its relationship to dynamic mode decomposition and resolvent analysis},
	volume = {847},
	issn = {0022-1120, 1469-7645},
	url = {https://www.cambridge.org/core/product/identifier/S0022112018002835/type/journal_article},
	doi = {10.1017/jfm.2018.283},
	language = {en},
	urldate = {2022-09-05},
	journal = {J. Fluid Mech.},
	author = {Towne, A. and Schmidt, O. T. and Colonius, T.},
	month = jul,
	year = {2018},
	pages = {821--867}
}

@article{choi_space--time_2019,
	title = {Space--time least-squares {P}etrov--{G}alerkin projection for nonlinear model reduction},
	volume = {41},
	issn = {1064-8275, 1095-7197},
	url = {https://epubs.siam.org/doi/10.1137/17M1120531},
	doi = {10.1137/17M1120531},
	language = {en},
	number = {1},
	urldate = {2025-04-14},
	journal = {SIAM J. Sci. Comput.},
	author = {Choi, Y. and Carlberg, Kevin},
	month = jan,
	year = {2019},
	pages = {A26--A58}
}

@article{leclercq_linear_2019,
	title = {Linear iterative method for closed-loop control of quasiperiodic flows},
	volume = {868},
	issn = {0022-1120, 1469-7645},
	url = {https://www.cambridge.org/core/journals/journal-of-fluid-mechanics/article/abs/linear-iterative-method-for-closedloop-control-of-quasiperiodic-flows/1A8A316A97FACCBB4BE2B3571F8326BA},
	doi = {10.1017/jfm.2019.112},
	language = {en},
	urldate = {2025-04-14},
	journal = {J. Fluid Mech.},
	author = {Leclercq, C. and Demourant, F. and Poussot-Vassal, C. and Sipp, D.},
	month = jun,
	year = {2019},
	pages = {26--65}
}

@article{lasagna_sensitivity_2020,
	title = {Sensitivity of long periodic orbits of chaotic systems},
	volume = {102},
	issn = {2470-0045, 2470-0053},
	url = {https://link.aps.org/doi/10.1103/PhysRevE.102.052220},
	doi = {10.1103/PhysRevE.102.052220},
	language = {en},
	number = {5},
	journal = {Phys. Rev. E},
	author = {Lasagna, D.},
	month = nov,
	year = {2020},
	pages = {052220}
}

@article{rubini_l_2020,
	title = {The $l_1$-based sparsification of energy interactions in unsteady lid-driven cavity flow},
	volume = {905},
	issn = {0022-1120, 1469-7645},
	url = {https://www.cambridge.org/core/product/identifier/S0022112020007077/type/journal_article},
	doi = {10.1017/jfm.2020.707},
	language = {en},
	journal = {J. Fluid Mech.},
	author = {Rubini, R. and Lasagna, D. and Da Ronch, A.},
	month = dec,
	year = {2020},
	pages = {A15}
}

@article{suri_capturing_2020,
	title = {Capturing {Turbulent} {Dynamics} and {Statistics} in {Experiments} with {Unstable} {Periodic} {Orbits}},
	volume = {125},
	issn = {0031-9007, 1079-7114},
	url = {https://link.aps.org/doi/10.1103/PhysRevLett.125.064501},
	doi = {10.1103/PhysRevLett.125.064501},
	language = {en},
	number = {6},
	urldate = {2024-09-18},
	journal = {Phys. Rev. Lett.},
	author = {Suri, B. and Kageorge, L. and Grigoriev, R. O. and Schatz, M. F.},
	month = aug,
	year = {2020},
	pages = {064501}
}

@article{ahmed_closures_2021,
	title = {On closures for reduced order models--{A} spectrum of first-principle to machine-learned avenues},
	volume = {33},
	issn = {1070-6631, 1089-7666},
	url = {https://aip.scitation.org/doi/10.1063/5.0061577},
	doi = {10.1063/5.0061577},
	language = {en},
	number = {9},
	urldate = {2022-11-28},
	journal = {Phys. Fluids},
	author = {Ahmed, S. E. and Pawar, S. and San, O. and Rasheed, A. and Iliescu, T. and Noack, B. R.},
	month = sep,
	year = {2021},
	pages = {091301}
}

@article{Barthel_Zhu_McKeon_2021,
	title={Closing the loop: nonlinear {T}aylor vortex flow through the lens of resolvent analysis},
	volume={924},
	DOI={10.1017/jfm.2021.623},
	journal={J. Fluid Mech.},
	author={Barthel, B. and Zhu, X. and McKeon, B.},
	year={2021},
	pages={A9}
}

@article{graham_exact_2021,
	title = {Exact {Coherent} {States} and the {Nonlinear} {Dynamics} of {Wall}-{Bounded} {Turbulent} {Flows}},
	volume = {53},
	issn = {0066-4189, 1545-4479},
	url = {https://www.annualreviews.org/doi/10.1146/annurev-fluid-051820-020223},
	doi = {10.1146/annurev-fluid-051820-020223},
	language = {en},
	number = {1},
	urldate = {2022-11-28},
	journal = {Annu. Rev. Fluid Mech.},
	author = {Graham, M. D. and Floryan, D.},
	month = jan,
	year = {2021},
	pages = {227--253}
}

@article{jin_energy_2021,
	title = {Energy transfer mechanisms and resolvent analysis in the cylinder wake},
	volume = {6},
	issn = {2469-990X},
	url = {https://link.aps.org/doi/10.1103/PhysRevFluids.6.024702},
	doi = {10.1103/PhysRevFluids.6.024702},
	language = {en},
	number = {2},
	journal = {Phys. Rev. Fluids},
	author = {Jin, B. and Symon, S. and Illingworth, S. J.},
	month = feb,
	year = {2021},
	pages = {024702}
}

@article{kaptanoglu_promoting_2021,
	title = {Promoting global stability in data-driven models of quadratic nonlinear dynamics},
	volume = {6},
	url = {https://link.aps.org/doi/10.1103/PhysRevFluids.6.094401},
	doi = {10.1103/PhysRevFluids.6.094401},
	number = {9},
	journal = {Phys. Rev. Fluids},
	author = {Kaptanoglu, A. A. and Callaham, J. L. and Aravkin, A. and Hansen, C. J. and Brunton, S. L.},
	month = sep,
	year = {2021},
	pages = {094401}
}

@inproceedings{towne_space-time_2021,
	address = {VIRTUAL EVENT},
	title = {Space-time {Galerkin} projection via {S}pectral {P}roper {O}rthogonal {D}ecomposition and resolvent modes},
	isbn = {978-1-62410-609-5},
	url = {https://arc.aiaa.org/doi/10.2514/6.2021-1676},
	doi = {10.2514/6.2021-1676},
	language = {en},
	booktitle = {{AIAA} {Scitech} 2021 {Forum}},
	publisher = {AIAA},
	author = {Towne, A.},
	month = jan,
	year = {2021}
}

@article{yalniz_coarse_2021,
	title = {Coarse {Graining} the {State} {Space} of a {Turbulent} {Flow} {Using} {Periodic} {Orbits}},
	volume = {126},
	issn = {0031-9007, 1079-7114},
	url = {https://link.aps.org/doi/10.1103/PhysRevLett.126.244502},
	doi = {10.1103/PhysRevLett.126.244502},
	language = {en},
	number = {24},
	urldate = {2024-09-18},
	journal = {Phys. Rev. Lett.},
	author = {Yalnız, G. and Hof, B. and Budanur, N. B.},
	month = jun,
	year = {2021},
	pages = {244502}
}

@article{azimi_constructing_2022,
	title = {Constructing periodic orbits of high-dimensional chaotic systems by an adjoint-based variational method},
	volume = {105},
	issn = {2470-0045, 2470-0053},
	url = {https://link.aps.org/doi/10.1103/PhysRevE.105.014217},
	doi = {10.1103/PhysRevE.105.014217},
	language = {en},
	number = {1},
	urldate = {2023-04-21},
	journal = {Phys. Rev. E},
	author = {Azimi, S. and Ashtari, O. and Schneider, T. M.},
	month = jan,
	year = {2022},
	pages = {014217}
}

@article{chua_khoo_sparse_2022,
	title = {A sparse optimal closure for a reduced-order model of wall-bounded turbulence},
	volume = {939},
	issn = {0022-1120},
	url = {https://www.cambridge.org/core/article/sparse-optimal-closure-for-a-reducedorder-model-of-wallbounded-turbulence/5371E960DDFD4893D319660CD19F8022},
	doi = {10.1017/jfm.2022.161},
	journal = {J. Fluid Mech.},
	author = {Khoo, Z. C. and Chan, C. H. and Hwang, Y.},
	year = {2022},
	pages = {A11}
}

@article{crowley_turbulence_2022,
	title = {Turbulence tracks recurrent solutions},
	volume = {119},
	issn = {0027-8424, 1091-6490},
	url = {https://pnas.org/doi/10.1073/pnas.2120665119},
	doi = {10.1073/pnas.2120665119},
	language = {en},
	number = {34},
	urldate = {2023-02-13},
	journal = {PNAS},
	author = {Crowley, C. J. and Pughe-Sanford, J. L. and Toler, W. and Krygier, M. C. and Grigoriev, R. O. and Schatz, M. F.},
	month = aug,
	year = {2022},
	pages = {e2120665119}
}

@article{karbasian_sensitivity_2022,
	title = {Sensitivity analysis of chaotic dynamical systems using a physics-constrained data-driven approach},
	volume = {34},
	issn = {1070-6631},
	url = {https://doi.org/10.1063/5.0076074},
	doi = {10.1063/5.0076074},
	number = {1},
	urldate = {2024-09-02},
	journal = {Phys. Fluids},
	author = {Karbasian, Hamid R. and Vermeire, Brian C.},
	month = jan,
	year = {2022},
	pages = {014101}
}

@article{parker_variational_2022,
	title = {Variational methods for finding periodic orbits in the incompressible {Navier}–{Stokes} equations},
	volume = {941},
	issn = {0022-1120, 1469-7645},
	url = {https://www.cambridge.org/core/product/identifier/S0022112022002993/type/journal_article},
	doi = {10.1017/jfm.2022.299},
	language = {en},
	urldate = {2023-04-21},
	journal = {J. Fluid Mech.},
	author = {Parker, J. P. and Schneider, T. M.},
	month = jun,
	year = {2022},
	pages = {A17}
}

@article{ashtari_identifying_2023,
	title = {Identifying invariant solutions of wall-bounded three-dimensional shear flows using robust adjoint-based variational techniques},
	volume = {977},
	issn = {0022-1120, 1469-7645},
	url = {https://www.cambridge.org/core/product/identifier/S0022112023009278/type/journal_article},
	doi = {10.1017/jfm.2023.927},
	language = {en},
	urldate = {2024-09-01},
	journal = {J. Fluid Mech.},
	author = {Ashtari, O. and Schneider, T. M.},
	month = dec,
	year = {2023},
	pages = {A7}
}

@misc{frame_linear_2024,
	title = {Linear model reduction using {SPOD} modes},
	url = {http://arxiv.org/abs/2407.03334},
	language = {en},
	urldate = {2024-07-10},
	publisher = {arXiv},
	author = {Frame, P. and Lin, C. and Schmidt, O. T. and Towne, A.},
	month = may,
	year = {2024},
	note = {arXiv:2407.03334 [cs, math]}
}

@misc{frame_space-time_2024,
	title = {Space-time model reduction in the frequency domain},
	url = {http://arxiv.org/abs/2411.13531},
	doi = {10.48550/arXiv.2411.13531},
	urldate = {2025-01-21},
	publisher = {arXiv},
	author = {Frame, P. and Towne, A.},
	month = nov,
	year = {2024},
	note = {arXiv:2411.13531 [math]},
}

@inproceedings{li_nonlinear_2024,
	address = {Montr\'eal, Canada},
	title = {Nonlinear frequency-domain reduced-order modelling of turbulent flows.},
	language = {en},
	booktitle = {13th {International} {Symposium} on {Turbulence} and {Shear} {Flow} {Phenomena} ({TSFP13})},
	author = {Li, X. and Lasagna, D.},
	month = jun,
	year = {2024}
}

@article{mccormack_multi-scale_2024,
	title = {Multi-scale invariant solutions in plane {Couette} flow: a reduced-order model approach},
	volume = {983},
	issn = {0022-1120, 1469-7645},
	shorttitle = {Multi-scale invariant solutions in plane {Couette} flow},
	url = {https://www.cambridge.org/core/product/identifier/S0022112024001083/type/journal_article},
	doi = {10.1017/jfm.2024.108},
	language = {en},
	journal = {J. Fluid Mech.},
	author = {McCormack, M. and Cavalieri, A. V. G. and Hwang, Y.},
	month = mar,
	year = {2024},
	pages = {A33}
}

@article{page_recurrent_2024,
	title = {Recurrent flow patterns as a basis for two-dimensional turbulence: {Predicting} statistics from structures},
	volume = {121},
	issn = {0027-8424, 1091-6490},
	shorttitle = {Recurrent flow patterns as a basis for two-dimensional turbulence},
	url = {https://pnas.org/doi/10.1073/pnas.2320007121},
	doi = {10.1073/pnas.2320007121},
	language = {en},
	number = {23},
	urldate = {2024-08-08},
	journal = {PNAS},
	author = {Page, J. and Norgaard, P. and Brenner, M. P. and Kerswell, R. R.},
	month = jun,
	year = {2024},
	pages = {e2320007121}
}

@article{tenderini_space-time_2024,
	title = {Space-{Time} {Reduced} {Basis} {Methods} for {Parametrized} {Unsteady} {Stokes} {Equations}},
	volume = {46},
	issn = {1064-8275},
	url = {https://epubs.siam.org/doi/abs/10.1137/22M1509114},
	doi = {10.1137/22M1509114},
	number = {1},
	urldate = {2025-04-14},
	journal = {SIAM J. Sci. Comput.},
	author = {Tenderini, R. and Mueller, N. and Deparis, S.},
	month = feb,
	year = {2024},
	pages = {B1--B32}
}

@article{burton_resolvent-based_2025,
	title = {Resolvent-based optimization for approximating the statistics of a chaotic {Lorenz} system},
	volume = {111},
	url = {https://link.aps.org/doi/10.1103/PhysRevE.111.025104},
	doi = {10.1103/PhysRevE.111.025104},
	number = {2},
	urldate = {2025-04-14},
	journal = {Phys. Rev. E},
	author = {Burton, T. and Symon, S. and Sharma, A. S. and Lasagna, D.},
	month = feb,
	year = {2025},
	pages = {025104}
}

\end{document}